\documentclass[sn-mathphys]{sn-jnl}

\jyear{2023}%

\theoremstyle{thmstyleone}%
\theoremstyle{thmstyletwo}%

\theoremstyle{thmstylethree}%

\newcommand{\Oii}{[O~{\sc ii}]}
\newcommand{\Oiia}{[O~{\sc ii}]$\lambda3727$} 
\newcommand{\Oiib}{[O~{\sc ii}]$\lambda3730$} 

\newcommand{\Neiiia}{[Ne~{\sc iii}]~$\lambda$3870}
\newcommand{\Neiiib}{[Ne~{\sc iii}]~$\lambda$3969}
\newcommand{\Oiii}{[O~{\sc iii}]}
\newcommand{\He}{H$\epsilon$}

\newcommand{\jwst}{\textit{JWST}}
\newcommand{\NIRSpec}{\textit{NIRSpec}}
\newcommand{\NIRCam}{\textit{NIRCam}}

\newcommand{\mnras}{Monthly Notices of the Royal Astronomical Society}
\newcommand{\nat}{Nature}
\newcommand{\apj}{The Astrophysical Journal}
\newcommand{\apjl}{The Astrophysical Journal Letters}
\newcommand{\apjs}{The Astrophysical Journal Supplements}
\newcommand{\aap}{Astronomy \& Astrophysics}
\newcommand{\aj}{The Astronomical Journal}
\newcommand{\rmxaa}{Revista Mexicana de Astronomia y Astrofisica}
\newcommand{\pasp}{Publications of the Astronomical Society of the Pacific}
\newcommand{\araa}{Annual Review of Astronomy and Astrophysics}

\begin{document}
\title{A massive interacting galaxy 510 million years after the Big Bang}{A massive interacting galaxy 510 million years after the Big Bang}



\author*[1,2]{Kristan Boyett}\email{kit.boyett@unimelb.edu.au}
\author*[1,2]{Michele Trenti}\email{michele.trenti@unimelb.edu.au}
\author[3]{Nicha Leethochawalit} 

\author[4]{Antonello Calabr\'o} 
\author[1,2,5]{Benjamin Metha} 

\author[5]{Guido Roberts-Borsani}
\author[1,2]{Nicol\'o Dalmasso} 
\author[6]{Lilan Yang} 
\author[4]{Paola Santini}

\author[5]{Tommaso Treu}
\author[7]{Tucker Jones}
\author[8,9]{Alaina Henry}

\author[10,11]{Charlotte A. Mason}
\author[12]{Takahiro Morishita}
\author[13]{Themiya Nanayakkara}

\author[9]{Namrata Roy}
\author[14,15,16]{Xin Wang}

\author[4]{Adriano Fontana}
\author[4]{Emiliano Merlin}
\author[4]{Marco Castellano}
\author[4]{Diego Paris}

\author[17,7]{Maru\v{s}a Brada\v{c}}
\author[5]{Matt Malkan}
\author[18]{Danilo Marchesini}
\author[4]{Sara Mascia}
\author[13]{Karl Glazebrook}
\author[4]{Laura Pentericci}
\author[19]{Eros Vanzella}
\author[20]{Benedetta Vulcani}


\affil*[1]{School of Physics, University of Melbourne, Parkville 3010, VIC, Australia}
\affil[2]{ARC Centre of Excellence for All Sky Astrophysics in 3 Dimensions (ASTRO 3D), Australia}
\affil[3]{National Astronomical Research Institute of Thailand (NARIT), Mae Rim, Chiang Mai, 50180, Thailand}
\affil[4]{INAF Osservatorio Astronomico di Roma, Via Frascati 33, 00078 Monteporzio Catone, Rome, Italy}
\affil[5]{Department of Physics and Astronomy, University of California, Los Angeles, 430 Portola Plaza, Los Angeles, CA 90095, USA}
\affil[6]{Kavli Institute for the Physics and Mathematics of the Universe, The University of Tokyo, Kashiwa, Japan 277-8583}
\affil[7]{Department of Physics and Astronomy, University of California Davis, 1 Shields Avenue, Davis, CA 95616, USA}
\affil[8]{Space Telescope Science Institute, 3700 San Martin Drive, Baltimore MD, 21218} 
\affil[9]{Center for Astrophysical Sciences, Department of Physics and Astronomy, Johns Hopkins University, Baltimore, MD, 21218}
\affil[10]{Cosmic Dawn Center (DAWN), Denmark}
\affil[11]{Niels Bohr Institute, University of Copenhagen, Jagtvej 128, DK-2200 Copenhagen N, Denmark}
\affil[12]{IPAC, California Institute of Technology, MC 314-6, 1200 E. California Boulevard, Pasadena, CA 91125, USA}
\affil[13]{Centre for Astrophysics and Supercomputing, Swinburne University of Technology, PO Box 218, Hawthorn, VIC 3122, Australia}
\affil[14]{School of Astronomy and Space Science, University of Chinese Academy of Sciences (UCAS), Beijing 100049, China}
\affil[15]{National Astronomical Observatories, Chinese Academy of Sciences, Beijing 100101, China}
\affil[16]{Institute for Frontiers in Astronomy and Astrophysics, Beijing Normal University,  Beijing 102206, China}
\affil[17]{University of Ljubljana, Department of Mathematics and Physics, Jadranska ulica 19, SI-1000 Ljubljana, Slovenia}
\affil[18]{Physics and Astronomy Department, Tufts University, 574 Boston Avenue, Medford, MA 02155, USA}
\affil[19]{INAF -- OAS, Osservatorio di Astrofisica e Scienza dello Spazio di Bologna, via Gobetti 93/3, I-40129 Bologna, Italy}
\affil[20]{INAF Osservatorio Astronomico di Padova, vicolo dell'Osservatorio 5, 35122 Padova, Italy}


\abstract{JWST observations spectroscopically confirmed the existence of galaxies as early as 300 million years after the Big Bang and with a higher number density than what was expected based on galaxy formation models and Hubble Space Telescope observations. Yet, the majority of sources confirmed spectroscopically so far in the first 500 million years have rest-frame UV-luminosities below the characteristic luminosity ($M_{UV}^*$), limiting the signal to noise ratio for investigating substructure. Here, we present a high-resolution spectroscopic and spatially resolved study of a bright ($M_{UV}=-21.66\pm0.03$, $\sim2M_{UV}^*$) galaxy at a redshift $z=9.3127\pm0.0002$  (510 million years after the Big Bang) with an estimated stellar mass of  $(1.6^{+0.5}_{-0.4})\times10^9~\mathrm{M_{\odot}}$, forming $19^{+5}_{-6}$ Solar masses per year and with a metallicity of about one tenth of Solar. The system has a morphology typically associated to two interacting galaxies, with a two-component main clump of very young stars (age less than $10$ million years) surrounded by an extended stellar population  ($120\pm20$ million years old, identified from modeling of the NIRSpec spectrum) and an elongated clumpy tidal tail. The observations acquired at high spectral resolution identify 
oxygen, neon and hydrogen emission lines, as well as the Lyman break, where there is evidence of substantial absorption of Ly$\alpha$. The \Oii~ doublet is resolved spectrally, enabling an estimate of the electron number density and ionization parameter of the interstellar medium and showing higher densities and ionization than in analogs at lower redshifts. 
We identify evidence of absorption lines (silicon, carbon and iron), with low confidence individual detections but signal-to-noise ratio larger than 6 when stacked. These absorption features suggest that Ly$\alpha$ is damped by the interstellar and circumgalactic medium. 
Our observations provide evidence of rapid and efficient build up of mass and metals in the immediate aftermath of the Big Bang through mergers, demonstrating that massive galaxies with several billion stars are in place at early times.} 


\maketitle


\section{Main} \label{sec:main}

The first generations of stars and galaxies in the Universe formed in physical conditions different to those of the modern Universe. In fact, gas is expected to have nearly primordial composition, with low levels of chemical enrichment and dust content (e.g., \cite{Torrey2019}). Gas cooling is further limited by the higher Cosmic Microwave Background radiation, possibly altering the characteristic fragmentation mass of protostellar clouds \citep{Smith2009}. In addition, early forming galaxies are expected to experience an elevated merger rate \citep{Fakhouri2008}, affecting their morphology and stellar populations.  

Hubble Space Telescope observations have been able to identify galaxy candidates at redshift $z\sim8-11$ ($\sim 600-400$ million years after the Big Bang; e.g. see \cite{Bouwens21}), and follow-up with the Spitzer Space Telescope provided evidence of relatively old stellar populations, suggesting that star formation started at $z>15$ \citep{Hashimoto2018,Roberts-Borsani20}. Yet, comparison to theoretical and numerical modeling has been restricted to number counts and luminosity functions by the limited angular resolution and dearth of spectroscopic data.  

With \jwst\ commencing science operations in July 2022, progress has been rapid and transformational. Already in the first Cycle of science programs, \jwst\ has built a convincing sample of $z>10$ candidates based on NIRCam photometry \citep{Castellano2022,naidu22,Donnan23,Harikane23,MorishitaStiavelli2022,Bouwens22, Adams23}, and a growing number of sources at $z\gtrsim 8$ are being confirmed spectroscopically with NIRSpec \citep{Morishita22, Williams22, Roberts-Borsani22b, Curtis-lake22, Tang23, Mascia23, Fujimoto23, Wang22, Cameron23, Bunker23, Hsiao23arXiv, Arrabal_Haro23a, Arrabal_Haro23b, Harikane23b}.
Surprisingly, the number density of (spectroscopically confirmed) high-redshift galaxies has been higher than expected by most models, in particular at the bright end of the luminosity function, possibly suggesting that we are missing key physical processes connected to the formation of first galaxies \citep{Mason2023,Wilkins23},  though observations do not suggest inconsistency with $\Lambda$CDM \citep{Harikane23b, Keller23, McCaffrey23}. 
The superior spectral and spatial resolution of \jwst\ in the near-infrared offers a path to investigate these initial findings through detailed stellar population studies in early galaxies. An initial spectroscopic census of \jwst\ galaxies above $z>7$ paints a picture of elevated ionization parameters, low metallicities, low dust content, high specific star formation rates (sSFR), and potentially higher ionizing radiation escape fraction ($f_{esc}$) than seen locally (e.g., \cite{Katz23, Curti23, Hsiao2022, Cameron23, Endsley22arxiv, Tang23, Sanders23}). Many of these galaxies show resemblance to the extreme interstellar medium (ISM) conditions observed in metal-poor actively star forming galaxies, both local (blueberries, greenpeas \citep{Cardamone09, Yang17, Henry15, Jaskot13}) and at moderate redshifts ($z\sim2$, \citep{Tang21, Sanders20, Sanders_2015}). The spatial resolution of \jwst\, also means the resolved stellar populations of galaxies in the epoch of reionization can be studied as well, revealing that these systems, once considered compact with HST, exhibit spatial variations in their physical properties \citep{Wang22b, Chen23, Gimenez-Arteaga22}. 

In this work we extend the frontier of detailed investigations of individual galaxy properties at very high redshift by reporting on imaging and spectroscopic observations of one of the brightest among the galaxy candidates at $z\gtrsim 9$ observed with \jwst, with flux $\sim 0.35~\mathrm{\mu Jy}$ in F444W (corresponding to $m_{AB}\sim 25.0$). These observations include 6-band NIRCam imaging (F115W, F150W, F200W, F277W, F356W, F444W, presented in Extended Data Table \ref{tab:photometry},) and NIRSpec high resolution ($R\sim2700$) Multi-object Spectroscopy (see~\ref{sec:obs}). The galaxy - which we call here Gz9p3 - was initially identified as a potential F105W or F125W dropout based on HST observations, and then confirmed as a NIRCam F115W drop-out with a probable $z\sim 9.45$ redshift \citep{Castellano22b}. 

The NIRSpec data we use are part of the GLASS-JWST program ERS-1324 \citep{TreuGlass22} centered on the foreground galaxy cluster Abell 2744, which is gravitationally lensing the high-redshift background sources. Gz9p3 is located in the outskirts of the cluster and relatively far away from high-magnification regions, with a best estimate of the lensing magnification $\mu=1.66\pm0.02$ based on \cite{Bergamini2022} (see~\ref{sec:lensing}). 

\begin{figure*}
    \centering
    \includegraphics[width=\textwidth]{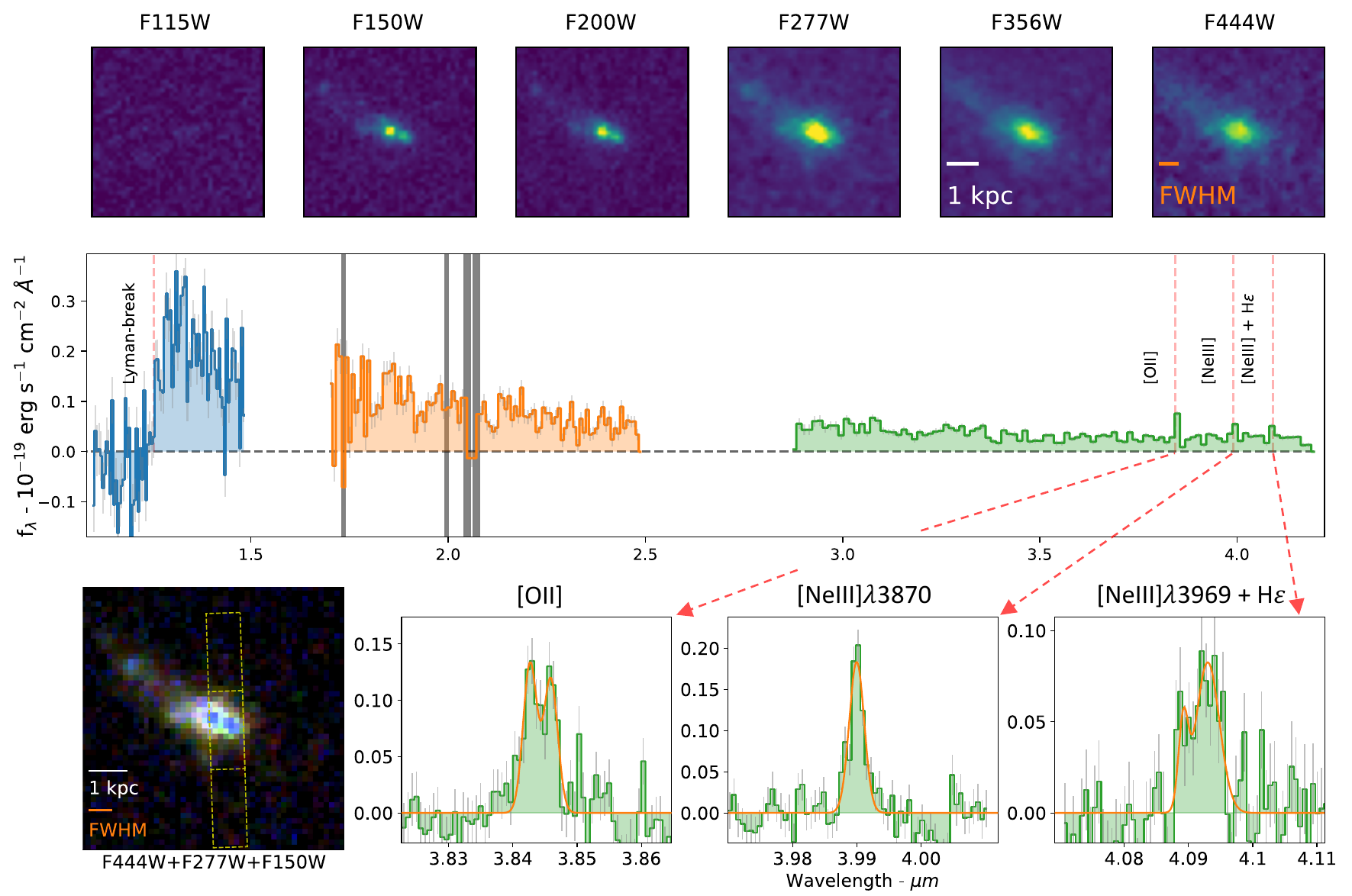}
    \caption{\textbf{\jwst\ NIRCam and NIRSpec observations of Galaxy GLASS ID:Gz9p3}. Top row: NIRCam direct imaging in broad-band filters. Second row: 1D standard extracted spectrum in units of $f_\lambda = 10^{-19} $erg s$^{-1} $cm$^{-2}$ \AA$^{-1}$ (with Npix=20 binning) from NIRSpec F100LP/G170H, F170LP/G235H $\&$ F290LP/G395H filter-dispersor configurations (from left-to-right in blue, orange and green). The observed frame location of the Lyman-break ($\lambda_{\rm rest}=1215.67$) and the \Oii, \Neiiia\ $\&$ \Neiiib+\He\ emission lines are overlaid as vertical dashed red lines for a $z_{\rm spec}=9.313$. The Solid grey regions mask the location of contaminating emission lines from  
    the dispersed spectrum of a galaxy falling in an open shutter of a separate quadrant of the NIRSpec MSA. Bottom left panel shows the color composite from NIRCam with NIRSpec slit positioning overlayed. The other three bottom panels are zoom in on a $\pm400$\AA\ region centered on each emission line complexes (\Oii, \Neiiia, \Neiiib+\He) with the continuum subtracted and the best-fit profile overlaid (again in units of $10^{-19}$ erg s$^{-1}$ cm$^{-2}$ \AA$^{-1}$).}
    \label{fig:spec}
\end{figure*}

The \jwst\ photometry for Gz9p3 is shown in the top panel of Figure \ref{fig:spec}. The source has a magnification-corrected apparent AB magnitude  $m_{AB}=25.56\pm0.06$ in F444W, which for the cosmology adopted in this paper (see~\ref{sec:cosmology}) corresponds to $M_{AB}=-21.77\pm0.06$ -- approximately 50\% brighter than the characteristic luminosity ($M^*$) of galaxies at this time \citep{Bouwens21}. The figure also includes a 1-D extraction of the NIRSpec spectrum in the middle panel, with spectral coverage in the $[1.1:4.5]~\mathrm{\mu m}$ range (with some gaps due to the target location relative to the edge of the instrument field of view as discussed in~\ref{sec:obs_spec}). 
We present the full 2D spectrum in Extended Data Figure \ref{fig:2d_spec}.
The NIRSpec Multi-shutter Assembly (MSA) was configured based on HST imaging and as such the shutter covers the main body of the galaxy as shown in the bottom left panel of the figure. The spectrum shows a clear continuum detection and four emission lines are detected at $[3.8:4.1]~\mathrm{\mu m}$, which we identify as (\Oii, \Neiiia, and \Neiiib\ blended with \He) for a source located at a redshift $z_{spec}=9.3127\pm0.0002$ (see ~\ref{sec:redshift}). In addition, the spectrum shows a Lyman break, and by modeling the stellar continuum as a step function around the break, we determine $z_{break} =9.35_{-0.05}^{+0.01}$, consistent with the redshift measurement from emission lines (shown in Extended Data Figure \ref{fig:lyman_break}, with further details in~\ref{sec:lyman_break_redshift}). 

\begin{figure*}
    \centering
    \includegraphics[width=\textwidth]{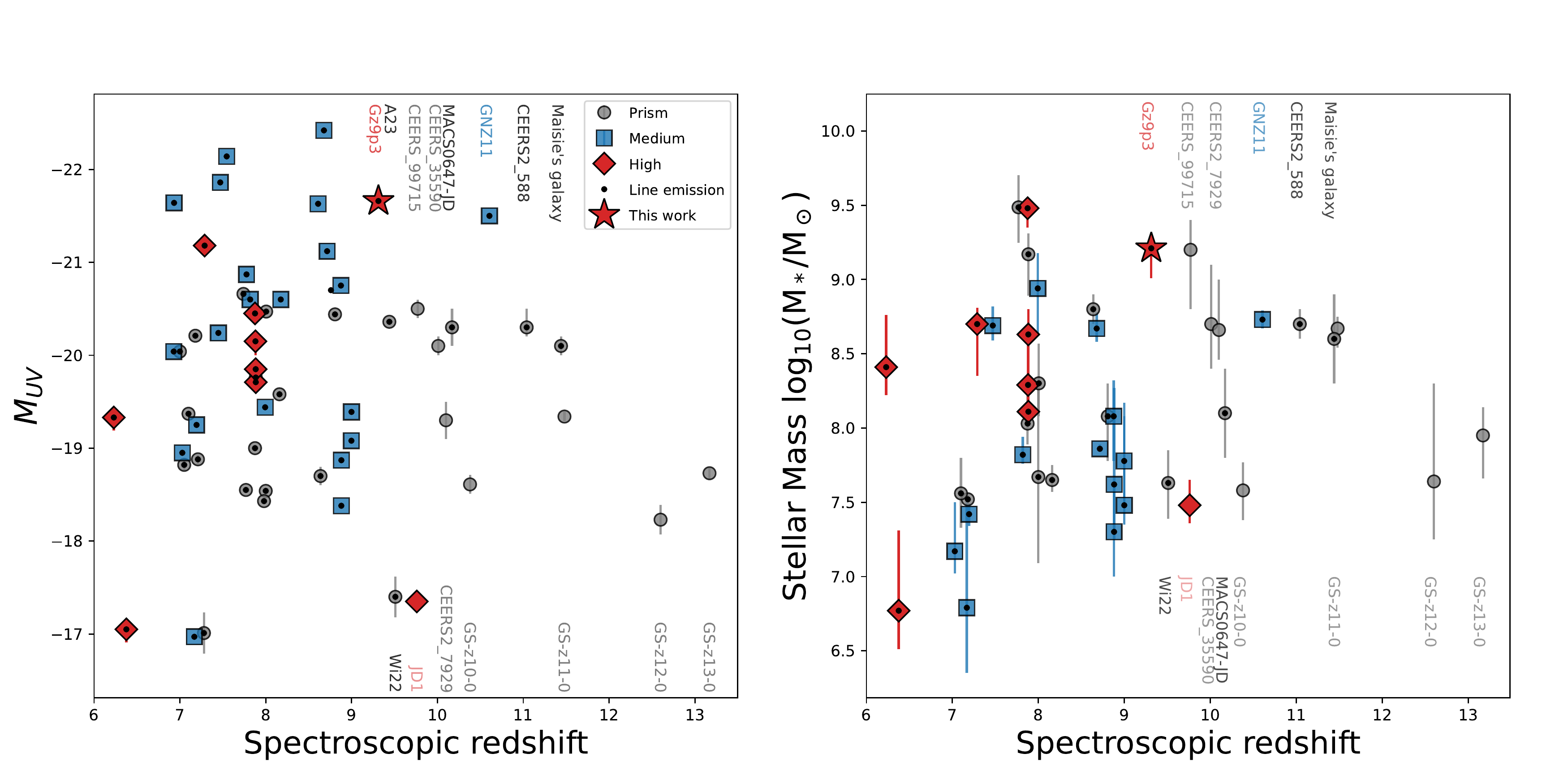}
    \caption{\textbf{Census of $M_{UV}$ and stellar mass in high-redshift galaxies.} Left: Absolute UV magnitude of galaxies spectroscopically confirmed with \jwst, corrected for lensing magnification where appropriate. Right: Stellar Mass (log$_{10}($M$_*/$M$_\odot)$) distribution of confirmed galaxies, corrected for magnification where appropriate. Error bars derive from the measured mass or $M_{UV}$ of individual galaxies.
    In both panels, the point shape represents the resolution mode of the NIRSpec spectroscopy (low-resolution prism: circles; Medium resolution: squares; High resolution: diamonds and star). A black central dot marker indicates the detection of emission lines. 
    Our target (red star with central dot) is one of the intrinsically brightest and most massive galaxies in the epoch of reionization among the current \jwst\ samples from 
    \cite{Williams22, Roberts-Borsani22b, Curtis-lake22, Tang23, Mascia23, Fujimoto23, Wang22, Cameron23, Bunker23, Hsiao23arXiv, Arrabal_Haro23a, Arrabal_Haro23b, Harikane23b}, and the highest and most massive at $z>9$. It is also one of the highest redshift galaxies with emission line detections and the highest redshift one observed in the high resolution mode. Stellar masses were taken as quoted from each study. 
    We note that we do not include MACS1149-JD1 \cite{Hashimoto18} as a spectroscopically confirmed galaxy (at z=9.11) due to significant uncertainty on its lensing magnification, and hence also on its intrinsic $M_{UV}$ and Stellar Mass. Additionally, we note that only a subset of the \cite{Tang23} sample have their masses reported, and masses aren't provided for \cite{Cameron23}.  We label all galaxies at $z>9$. In these labels A23 refers to galaxy ID: 10058975 for \cite{Cameron23} and Wi22 refers to \cite{Williams22}.}
    \label{fig:Muv_redshift}
\end{figure*}

The galaxy properties are derived from Spectral Energy Distribution (SED) modeling using both broad-band photometry and the 1-D spectrum as input (see Extended Data Figure \ref{fig:spec_phot}, with further details in ~\ref{sec:SED}). The results are summarized in Table~\ref{tab:properties}. Based on photometry of the whole galaxy (Kron fluxes, see \cite{Paris23} and~\ref{sec:SED}), the modeling returns a magnification-corrected stellar mass of  $\log_{10}($M$_*/$M$_\odot)=9.2^{+0.1}_{-0.2}$ and $M_{UV}=-21.66\pm0.03$. (These are statistical uncertainties from photometric errors only). This makes Gz9p3 one of the most massive and intrinsically brightest galaxies confirmed in the epoch of reionization, and the brightest and most massive at $z>9$ (see Figure \ref{fig:Muv_redshift}  for a census for high-redshift spectroscopically confirmed galaxies).
Even when compared against photometric galaxy candidates \cite{Labbe22}, Gz9p3 has one of largest masses known within the first 750Myr since the Big Bang. The SED modeling from the spectrum is restricted to the main region of the galaxy where the shutter was placed, and returns a stellar mass consistent with the photometric estimate (see~\ref{sec:spec_interp}). Both SED modeling approaches (photometry and spectrum fitting) also identify substantial ongoing star formation  ($9-19~\mathrm{M_{\odot}yr^{-1}}$), and limited evidence of dust due to the blue spectral slope $\beta$ in the UV ($-2.2 \lesssim \beta \lesssim -1.9$), with robust results over a range of assumed star formation histories. Interestingly, the spectrum-based modeling shows evidence for older stellar populations in the central region of the galaxy (age $120\pm 20$ Myr), indicating that star formation started as early as $z\gtrsim 15$ to produce the average age observed (see~\ref{sec:spec_interp}). In contrast, modeling based on photometry infers younger ages, with integrated-light fits giving an age of $25^{+15}_{-12}$ Myr and spatially resolved analysis identifying regions with ages $<10$ Myr (see~\ref{sec:SED_regions} and Extended Data Table~\ref{tab:region_results}).  

The detection of rest-optical emission lines provides a window into the interstellar medium conditions in the galaxy by resolving the \Oii\ doublet thanks to our high spectral resolution (Figure \ref{fig:spec}), measuring a line ratio of $0.94^{+0.14}_{-0.18}$ (see~\ref{sec:spec_interp}). The relative strength of these low-ionization lines is sensitive to the electron number density $n_e$ \citep{Sanders16b}, leading to a measurement of $n_e=590^{+570}_{-250}cm^{-3}$. This is marginally higher (at $\gtrsim 1\sigma$) than the median values of $n_e=225$cm$^{-3}$ seen in galaxies at $z=2.3$, and $n_e=26$cm$^{-3}$ seen in local galaxies \citep{Sanders16b}, qualitatively following the trend of $n_e$ increasing with redshift as reported in that study.
From the spectrum we determine a Ne3O2 (\Neiiia/\Oii) ratio of $0.81\pm0.09$. As these two lines are close in wavelength, the ratio is insensitive to dust reddening. The measurement is higher than what is typically seen at lower redshift, indicating a high ionization parameter of log $U=-2.13\pm0.05$ based on \citep{Witstok21}, and a low metallicity of $12+\log($O/H$) = 7.6 \pm 0.5$ (depending on which one among the low-$z$ Ne3O2 calibrations is used and including systematic uncertainties \cite{Bian18, Shi07, Maiolino_2008, Jones15}; see~\ref{sec:metallicity} for further details). 
Together, these conditions indicate a sub-solar ($Z\lesssim 0.1Z_\odot$) metal-poor interstellar medium, with a high electron density and ionization parameter, exhibiting similar properties to other galaxies spectroscopically confirmed at $z_{spec}>8$  \citep{Katz23, Curti23, Mascia23, Hsiao2022, Fujimoto23, Sanders23, Cameron23}. The ISM conditions are consistent with expectations from the young stellar ages derived from SED fitting, providing a self-consistent picture of the stellar populations and their surrounding gas.

\begin{figure*}
    \centering
    \includegraphics[width=\textwidth]{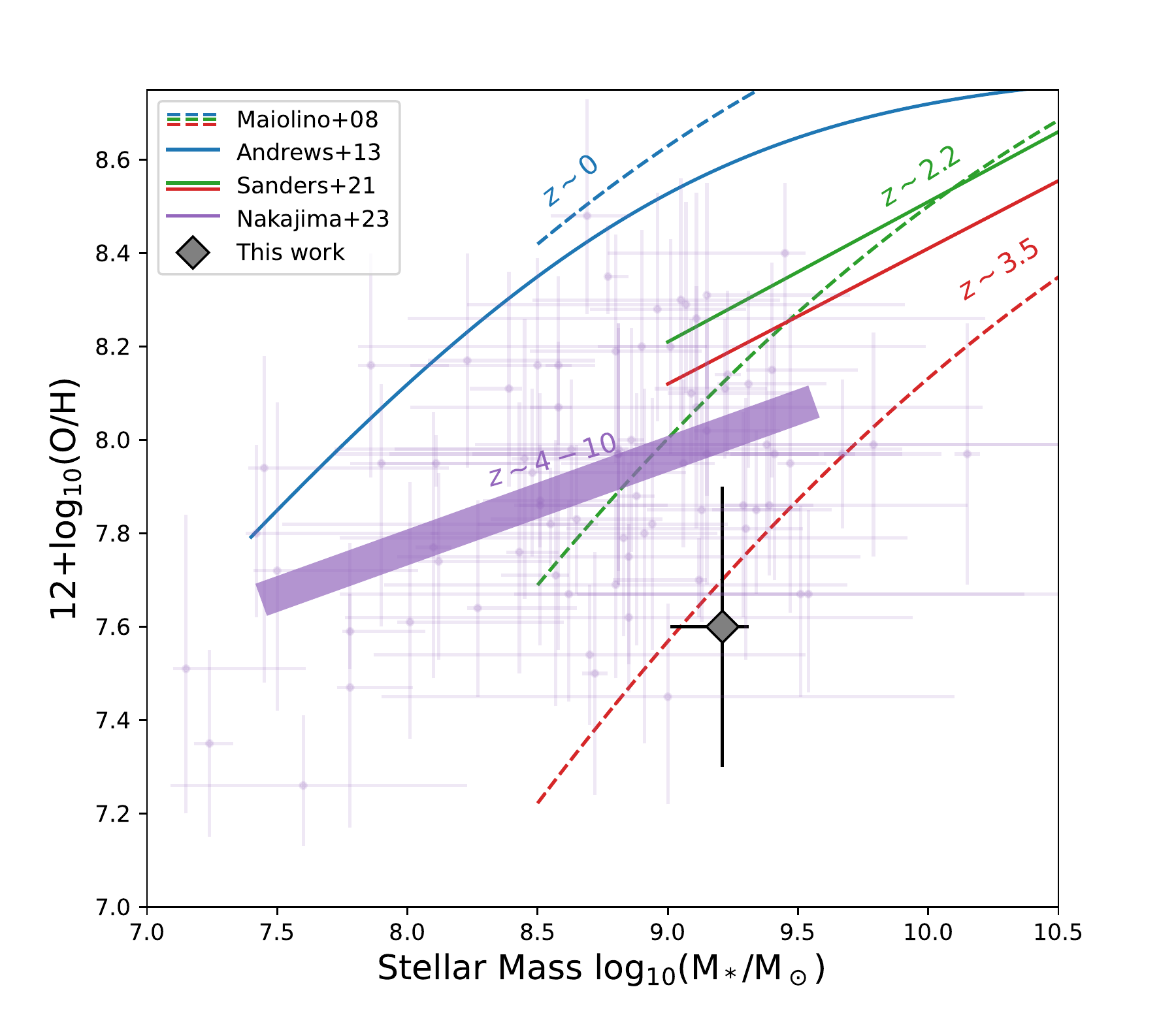}
    \caption{\textbf{Location of Gz9p3 on the Mass metallicity relation.} Gz9p3 is shown in black, where the error bar presents the random and systematic uncertainty in the metallicity, based on Ne3O2 diagnostic calibrations from \cite{Bian18, Shi07, Maiolino_2008, Nakajima22, Jones15}. The figure includes a comparison to mass metallicity relations covering 4 redshift epochs: $z\sim0$ (blue) from \cite{Andrews13, Maiolino_2008}, $z\sim2.2$ and $z\sim3.5$ (green and red) from \cite{Sanders21, Maiolino_2008} and $z=4-10$ (purple) from \cite{Nakajima23}. We additionally show the \jwst\ $z=4-10$ galaxies from \cite{Nakajima23} in purple, where the error bars derive from the mass and metallicity calculation for each individual galaxy.}
    \label{fig:mass_metal}
\end{figure*}

In Figure \ref{fig:mass_metal}, the large stellar mass and low oxygen abundance place Gz9p3 below the mass-metallicity relations for $z=4-9$ derived by  \cite{Nakajima23}, and marginally below the relation for $z=2-4$ galaxies from \cite{Sanders21, Maiolino_2008}, even though systematic uncertainty may affect the robustness of these conclusions (see~\ref{sec:metallicity}). The offset suggests Gz9p3 has a high gas fraction and potentially high accretion rates of pristine gas. This is qualitatively consistent with expectations from theoretical and numerical modeling of galaxies at these early times, given the short assembly times of their dark-matter halos \cite{Mason2015}. 

Further insight into the stellar populations of Gz9p3 is offered by the detection -  albeit at low confidence - of UV absorption lines, shown in Extended Data Figure \ref{fig:uv_absorption}. The spectrum shows a flux deficit associated with the SiII$\lambda$1260, CII$\lambda$1335 and FeII$\lambda$2344 lines at a $>4\sigma$ significance individually, and at  $>6\sigma$ when the lines are stacked together (see~\ref{sec:absorption_lines}). The detection of metal absorption features reinforces the evidence from emission lines that the gas within the galaxy is not pristine and has been enriched by the older stellar population, again providing a consistent interpretation of the relatively evolved stellar ages inferred from SED modeling for the core of the galaxy where the slit is placed. The multi-component absorption profile hinted by the data suggests turbulence within the absorbing gas, as expected from models of star formation \citep{Padoan02,Krumholz05}, and/or from bulk gaseous flows that are associated to a merging or interacting system. However, higher signal to noise observations are required to quantify the robustness and confidence of these interpretations.

Interestingly, not only is there no evidence of Ly$\alpha$ emission, with a stringent limit on the equivalent width from Table~\ref{tab:properties}, but also the stellar continuum at $1216$\AA$< \lambda_{\rm rest} < 1240$\AA\ (redward of the Lyman break) shows a deficit. Measurements over $\Delta \lambda_{\rm rest}=4$\AA\ and $24$\AA\ windows show a $80\%$ and $40\%$ deficit at a $5.7\sigma$ and $6.5\sigma$ significance, respectively (see~\ref{sec:lyman_break}). The softening of the spectral break supports the presence of absorption from Ly$\alpha$ damping wings (seen in the spectra of many $z>9$ galaxies, \citep{Umeda23arXiv, Heintz23arXiv, Hsiao23arXiv}). One interpretation is that the damping is due to absorption by the intergalactic medium \citep{Miralda_escude98}, which would indicate that the galaxy does not reside within a large ionized bubble. Such a scenario falls in line with the expected transmission due to damping wing in a neutral IGM from \cite{Mason20}, with a predicted flux of $\sim0.3\times$ continuum and $\sim0.8\times$ continuum at 1000 km\,s$^{-1}$ ($\sim4$\AA) and 6000 km\,s$^{-1}$ ($\sim24$\AA) respectively. However, it is difficult to reconcile the lack of a large ionizing bubble with the high stellar mass and presence of relatively old ($>100$ Myr) stars, especially because Ly$\alpha$ emission is detected in galaxies at higher redshift with lower star formation rates and stellar masses such as GN-z11 \cite{Bunker23}. An alternative interpretation is that the interstellar and circumgalactic medium in Gz9p3 is primarily responsible for the lack of Ly$\alpha$ emission in the spectrum, irrespective of the IGM conditions. This scenario is supported by the detection of the SiII$\lambda$1260 and CII$\lambda$1335 absorption features with a rest-frame equivalent width of $(-3.7\pm 0.8)$\AA. In fact, it has been shown that their strength correlates with damping of Ly$\alpha$ and results in Ly$\alpha$ absorption when the equivalent width of low-ionization interstellar metal lines is $\lesssim -2$\AA\ 
\cite{Jones2012,Leethochawalit2016,Pahl2020}. Also, the presence of strong low-ionization ISM line absorption and the stringent upper limit on CIII] emission suggest that the galaxy is unlikely to be a Lyman continuum leaker \cite{Lopez2022}.

\begin{figure*}
    \centering
    \includegraphics[width=\textwidth]{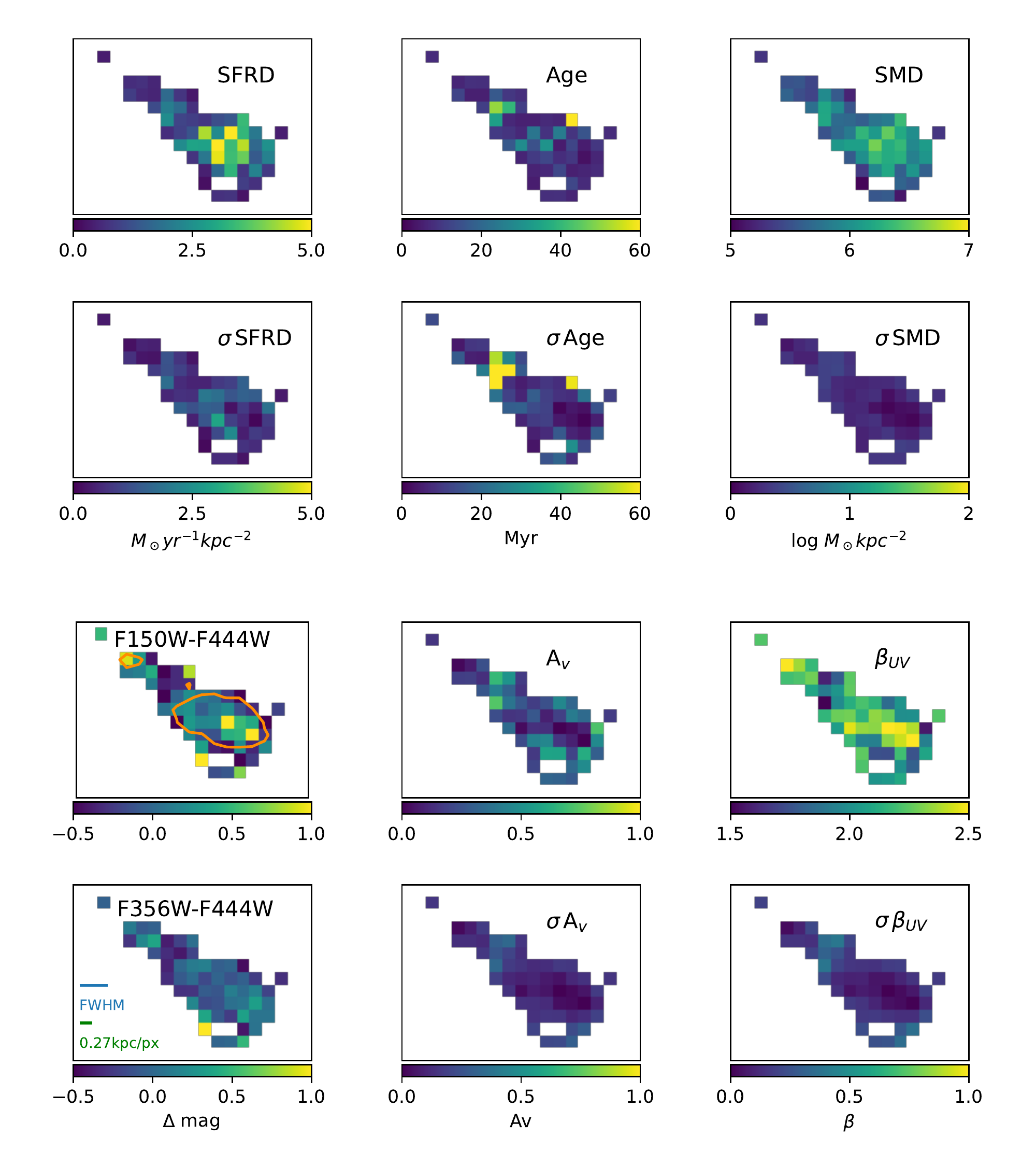}
    \caption{\textbf{2D color and physical parameter distribution of Gz9p3.} Properties inferred from photometric spectral energy distribution fitting (NIRCam pixels matched to F444W and binned 2x2), with an observed frame resolution of 0.27kpc/pixel (0.21kpc/pixel in the image plane). 
    From left to right: the top and second row present the star formation rate surface density, stellar age, and stellar mass surface density. The third and bottom row present the color, visual extinction and UV $\beta$ slope.  (Upper panels: Median value, Lower panels: Uncertainty based on 16$^{th}$ and 84$^{th}$ percentiles).
    The F150W $10\sigma$ contour is presented in orange in the F150W-F444W panel and the FWHM and pixel-scale are shown in the F356W-F444W panel.}
    \label{fig:maps}
\end{figure*}

In addition to providing detailed spectral insight on the stellar populations within the shutter aperture, the NIRSpec observations allow us to fix the redshift of Gz9p3 for photometric pixel-by-pixel modeling. This allows us to investigate spatial variations across the galaxy, following an approach similar to the one adopted by \cite{Gimenez-Arteaga22} at lower redshift, to create a 2D distribution of the galaxy's physical properties (see~\ref{sec:spatial_resolved_SEDs}). The analysis is carried out with BAGPIPES \cite{Carnall18} to determine the following maps of resolved properties, shown in Figure ~\ref{fig:maps}: stellar mass surface density (SMD);  100Myr-time averaged star formation rate surface density (SFRD); mass-weighted stellar age; visual extinction (Av); and UV $\beta$ slope (where $f_\lambda \propto \lambda^{-\beta}$).
The native pixel resolution in F444W is $0.031\mathrm{"/pixel}$ (with a FWHM F444W $\sim4.5$ native pixels, or $\sim0.14"$), but for our analysis we bin pixels $2\times2$ to improve the signal-to-noise ratio per pixel, generating the maps at a 0.27kpc/pixel resolution in the observed frame. This corresponds to 0.21kpc/pixel in the image plane after accounting for gravitational magnification. The galaxy shows a morphology comprised of an elongated tail and a central body, which in F444W appears as a single core.
The peak of the star forming activity within the stellar population is situated in the central core and this traces the distribution of stellar mass. This central region of active star formation exhibits young stellar population ages ($<50Myr$), as does the clump within the tail, whilst the surrounding regions on the outskirts of the central body show older populations, albeit with larger uncertainties ($50 \lesssim t_{\rm age}$/Myr $\lesssim 125$). The galaxy shows blue $\beta$ slopes throughout and relatively low visual extinction, which implies low dust content (see~\ref{sec:spatial_resolved_SEDs}). 

\begin{figure*}
    \centering
    \includegraphics[width=\textwidth]{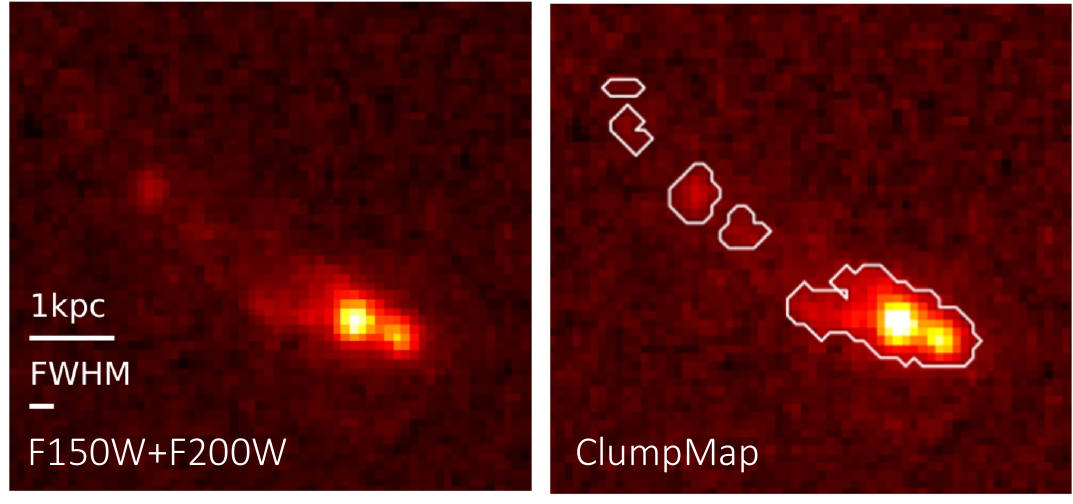}
    \caption{\textbf{Morphology of Gz9p3.} F150W+F200W direct image at a 20mas/px resolution in both panels. Left: direct imaging shows a double core within the central region and an elongated clumpy structure. Right: Overlaid Clump-map, showing 4 clumps detected within the tail of the system with a clumpiness parameter $c=0.56$.}
    \label{fig:clumpy}
\end{figure*}

The spatially resolved modeling is suggestive of an interacting system undergoing (or having recently undergone) a major merger. To further investigate this scenario, we analyze photometry at rest-frame UV wavelengths, using a combined F150W+F200W image drizzled at 20 mas/pixel, presented in Figure \ref{fig:clumpy}.
The data clearly show two distinct cores in the main region of the galaxy, and several components in the tail identified thorough a clump-finding algorithm (see~\ref{sec:merging}). The morphology of Gz9p3 is described by morphological parameters that indicate the galaxy is a merger (Gini=0.61, M20=-1.29, A=0.35; see~\cite{Lotz08}). 
\clearpage

Informed by the morphological analysis, we repeat the spatially resolved SED modeling by placing apertures over different stellar populations shown in Extended Data Figure \ref{fig:regions} (innermost region, a surrounding annulus and the tail), clearly seeing a distinction between active regions of star formation and an underlying older stellar population, as expected from the merging scenario (see~\ref{sec:SED_regions}). 

The combined spectroscopic and imaging data paint a picture of a very bright and relatively massive interacting system just 0.5 Gyr after the Big Bang, raising the question of how likely such \jwst\ observations should be. Figures~\ref{fig:Muv_redshift}-\ref{fig:mass_metal} hint that the system could be an outlier. To quantify expectations we consider both analytical modeling of early galaxy formation and comparison to cosmological hydrodynamical simulations. We find that while the likelihood of capturing an interacting system is relatively high, i.e. $\sim 20\%$ for a major merger, the stellar mass of Gz9p3 is higher than expected (see~\ref{sec:theory}-\ref{sec:simulations}). This would indicate either the system is hosted in a very rare dark matter halo for that epoch, that we serendipitously observed, or more likely that the current recipes for star formation are missing some key ingredients at early times\cite{Haslbauer22}. 
The latter interpretation would be consistent with the excess of sources identified by \jwst\ at $z>10$ through imaging programs \cite{Castellano22b} and with the high numbers of massive red galaxy candidates found at $z\sim 7.5-9$ \cite{Labbe22}. 
All these aspects make Gz9p3 an excellent target for further spectroscopic investigations, in particular through the Integral Field Unit mode on NIRSpec, that would shed further light on the kinematics of the system and on the complex interplay between assembly of dark matter halos, star formation and physical conditions in the interstellar, circumgalactic and intergalactic media at very early times.

\clearpage
\begin{table}
\centering
\begin{tabular}{cc}
Property   & Observed Value                             \\\hline \hline
RA [Deg]  & 3.617193\\
DEC [Deg] & -30.4255352\\
$z_{spec}^a$   & $9.3127\pm0.0002$\\
$\mu$  & $1.66\pm0.02$ \\
\hline \hline
\multicolumn{2}{c}{Line flux  [$10^{-19}$erg s$^{-1}$cm$^{-2}$]} \\
\hline
Ly$\alpha$ & $<2.64^\dag$\\
CIII]$\lambda1908$ & $<1.45^\dag$\\
MgII$\,\lambda2804$ & $<1.07^\dag$\\
\Oii\ & $6.7^{+0.4}_{-0.5}$ \\
\Neiiia\ &  $5.4^{+0.6}_{-0.5}$ \\
(\Neiiib\ + H$\epsilon$)  & $4.5^{+0.6}_{-0.4}$ \\
\Neiiib\  & $1.1^{+1.7}_{-0.9}$ \\ 
H$\epsilon$  & $3.4^{+1.1}_{-1.5}$ \\
\hline \hline
\multicolumn{2}{c}{Line EW$_{\rm rest}$ [\AA]} \\
\hline
Ly$\alpha$ & $<7.6^\dag$\\
CIII]$\lambda1908$ & $<1.0^\dag$\\
MgII$\,\lambda2804$ & $<1.9^\dag$\\
\Oii\ & $25.8^{+1.5}_{-1.9}$ \\
\Neiiia\  &  $21.4^{+2.4}_{-2.0}$ \\
(\Neiiib\ + H$\epsilon$)& $18.0^{+2.4}_{-1.6}$ \\
\Neiiib\  & $4.4^{+6.7}_{-3.6}$ \\ 
H$\epsilon$  & $13.5^{+4.4}_{-6.0}$ \\
\hline \hline
\multicolumn{2}{c}{Full photometry SED fit$^b$}\\
\hline
Stellar Mass [log$_{10}($M$_*/$M$_\odot)$] & $9.2^{+0.1}_{-0.2}$\\
SFR$^\dag$  [M$_\odot yr^{-1}$] &  $19^{+5}_{-6}$ \\
Stellar Age [Myr] &  $25^{+15}_{-12}$\\
$\beta$  & $-1.94^{+0.05}_{-0.06}$ \\
$M_{UV, SED}$ [AB\_mag] & $-21.66\pm0.03$ \\
\hline \hline
\multicolumn{2}{c}{Spectrum+photometry SED fit of main component$^c$}\\
\hline
Stellar Mass [log$_{10}($M$_*/$M$_\odot)$]& $9.15\pm0.04$\\
SFR$^\dag$  [M$_\odot yr^{-1}$] & $9.1\pm0.6$ \\
Stellar Age  [Myr] &  $120\pm20$\\
$\beta$  & $-2.23\pm0.04$ \\
$M_{UV, SED}$ [AB\_mag] & $-20.92\pm0.02$
\end{tabular}%
\caption{Physical properties for galaxy Gz9p3. 
$^\dag$$1\sigma$ upper limit. Line fluxes are not corrected for dust extinction or slit loss. 
$^b$ Photometry-only SED properties for the full system,  corrected where appropriate for magnification.
$^c$ Spectrum+Photometry SED properties for our main region aperture (see Extended Data Figure \ref{fig:regions}), corrected for magnification and adopting slit-losses based on aperture photometry.
$^\dag$ The SFR is taken as the 100Myr average from the BAGPIPES SFH model.
}
\label{tab:properties}
\end{table}


\section{Methods} \label{sec:methods}

\subsection{Cosmology and conventions}\label{sec:cosmology}

Where applicable, we use a standard $\Lambda$CDM cosmology with parameters \textit{H}$_{0}=70$ km\,s$^{-1}$\,Mpc$^{-1}$, 
$\Omega_{m}=$0.3, and $\Omega_{\wedge}=$0.7. All magnitudes are in the AB system \citep{Oke83}. Throughout this paper we refer to quantities as \lq\lq observed" or \lq\lq intrinsic" depending on whether the best available estimate of any gravitational lensing magnification correction has been applied. All wavelengths for emission lines are quoted as vacuum wavelengths.

\subsection{Observations}\label{sec:obs}

This work is based on \jwst/NIRCam \citep{rieke23} direct imaging and \jwst/NIRSpec \citep{jakobsen22,ferruit22} high resolution multi-object spectroscopy as part of the GLASS-JWST survey (ERS 1324, PI Treu; \cite{TreuGlass22}) and DDT program 2756 (PI Wenlei Chen).
We refer the reader to \cite{Roberts-Borsani2022, Paris23} for details of the observations and reduction strategy for NIRCam and to \cite{Morishita22} for NIRSpec.

As part of the GLASS \jwst/NIRSpec observations, over 100 galaxies were assigned to slit apertures. Our target (GLASS ID: 10003, indicated as Gz9p3 hereafter) was included as a candidate, based on HST ACS/WFC3 photometry \citep{Castellano2016} which identified it as a potential F105W/F125W drop-out. This placed the object at a redshift $z_{phot}>8$. Without longer wavelength imaging at the spatial resolution needed to deblend near-proximity neighbors, the validity of the single broadband filter detection in HST/WFC3 F160W could not be confirmed and physical properties could not be determined. 

The photometric redshift solution has since been refined to $z_{phot}=9.45$ based on NIRCam observations \citep[ID: DHZ1 in][]{Castellano22b}. This object was one of 7 $z\sim10$ galaxies identified in the ABELL 2744 cluster region, suggesting an apparent over-density at this epoch in the field \citep[see][]{Castellano22b} and reaffirming the priority for spectroscopic follow-up. \jwst\ NIRSpec observations now provide the spectral coverage needed to investigate this object further.

\subsubsection{High-resolution spectroscopy}\label{sec:obs_spec}
We obtain high-resolution ($R\sim\frac{\Delta\lambda}{\lambda}=2700$) \jwst/NIRSpec Multi-Object Spectroscopy of this target using the F100LP/G140H, F170LP/G235H and F290LP/G395H filter-disperser combinations.
These spectroscopy configurations combined with the location of the object on the detector provide wavelength coverage from $1.00-1.48\mu m$, $1.70-2.49\mu m$ and $2.88-4.19\mu m$, respectively (970-4060\AA\ in the rest-frame, see $z_{spec}$ discussion in \ref{sec:redshift}). The exposure time for each configuration is 4.9 hours. 
We note that while this work was undergoing peer-review and after an initial version of this manuscript had been posted on arXiv, observations of Gz9p3 using NIRSpec in the prism mode have been conducted by the UNCOVER team (\cite{Bezanson22, Fujimoto23b}) to extend the observed wavelength coverage out to $\sim5.2\mu m$, additionally detecting \Oiii+H$\beta$.
 
In Figure \ref{fig:spec}, we overlay the location of the NIRSpec open shutters on the RGB direct image. The open shutters are placed over the main component but do not capture the total flux from the galaxy and therefore suffer from some slit losses. 
Due to the extended morphology and potential change in stellar populations across the galaxy, we do not correct the slit-losses to match the total photometry of the galaxy. Instead we adopt a slit-loss correction based only on the photometry of the main component (where the slit is placed), as part of our SED spectrum+photometry analysis of this region (see  \ref{sec:spec_interp}).
Therefore, for measurement of physical properties for the whole galaxy (e.g., Mass, SFR in \ref{sec:SED}), we rely on the photometric SED fitting (see \cite{Paris23}), and use the spectrum SED fitting as consistency check and for investigation of the properties of the core of the system. 

A detailed discussion of how the data were reduced is given in \cite{Morishita22}. Briefly, we use the official STScI \jwst\ pipeline (ver.1.8.2)
for Level 1 data products, and the
\texttt{msaexp}
Python package for Level 2 and 3 data products. The orientation of the open-shutter configuration relative to Gz9p3 means the core of the galaxy is well-contained within the middle shutter, with the adjacent open shutters unoccupied, and so only sample the background. This orientation allows the adjacent open-shutters to be used for background subtraction and will not suffer from auto-subtraction.
We extract the 1D spectrum following \cite{Morishita22} and use the \texttt{msaexp} package to optimise our extraction using an inverse-variance weighted kernel. This kernel is derived by summing the 2D spectrum along the dispersion axis and fitting the resulting signal along the spatial axis with a Gaussian profile ($\sigma\sim0.4"$). This is then used to extract the 1D spectrum along the dispersion axis.

\begin{figure*}
    \renewcommand\thefigure{1}
    \renewcommand{\figurename}{Extended Data Figure.}
    \centering
    \includegraphics[width=\textwidth]{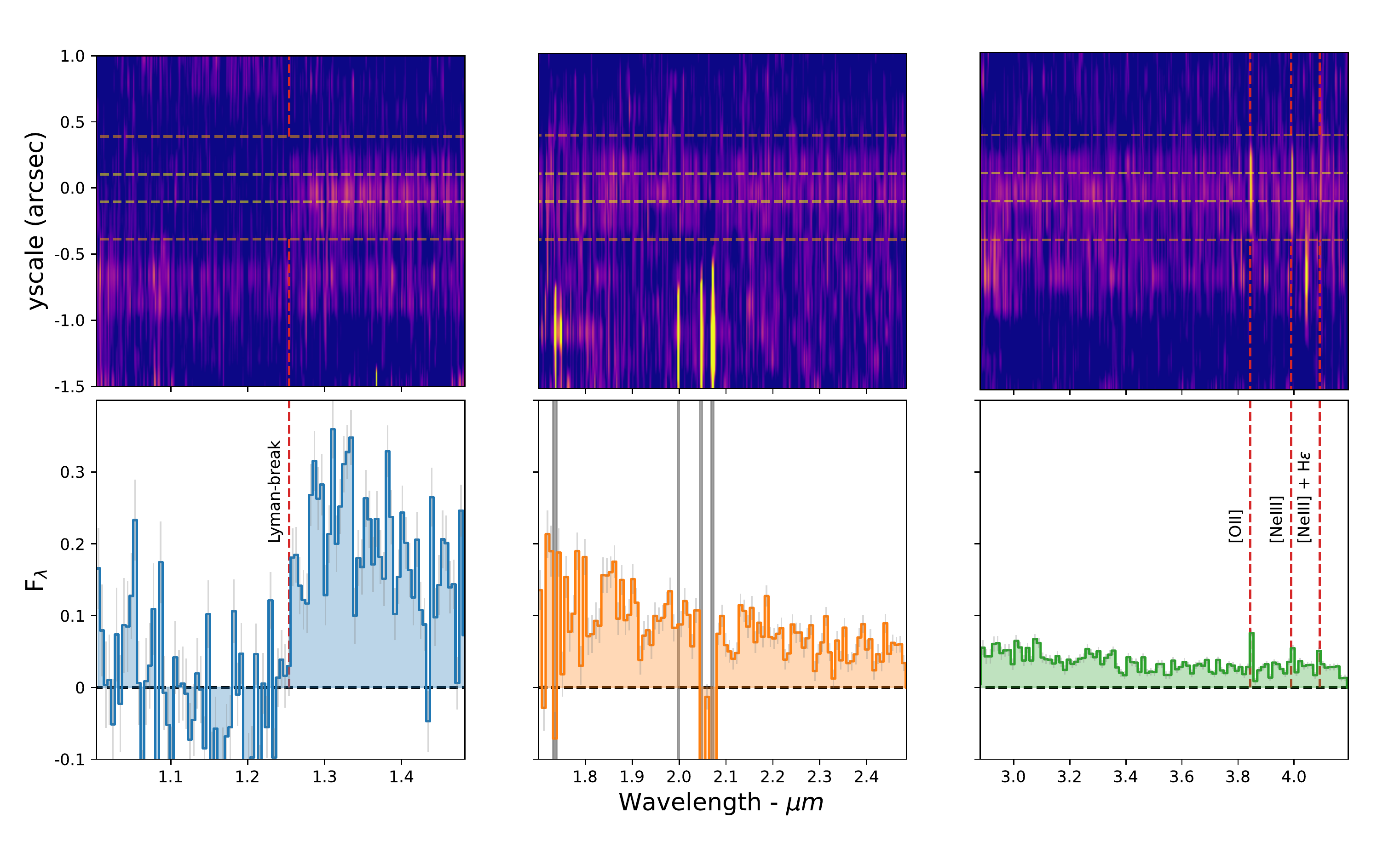}
    \caption{\textbf{NIRSpec 2D spectrum of Gz9p3.} Top: 2D observed-frame high resolution $R\sim2700$ spectrum in the f100lp/g140h, f170lp/g235h and f290lp/g395h filter-disperser configurations. Orange and white horizontal lines show the $1\sigma$ extraction trace for the optimal and narrow kernels. In the 2D extraction, the relative proximity of the dispersed light from other sources can be seen. These spectra are not associated with an additional source within our shutter but rather with targets in separate shutters in the NIRSpec MSA with a similar row number. The narrow kernel is introduced to ensure no contamination from the dispersed light of close proximity spectra is included for the fitting of the Lyman break. Bottom: the 1D extracted spectrum (as in Figure \ref{fig:spec}). We mark the location of the Lyman-break and detected emission lines in red in both the 1D and 2D spectra. The wavelength region contaminated by the additional source is marked in grey in the 1D.}  
    \label{fig:2d_spec}
\end{figure*}

We present the full 2D spectrum in Extended Data Figure \ref{fig:2d_spec} and the extracted 1D spectrum in Figure \ref{fig:spec}. Within the 2D image there is clear emission from an additional source below the trace associated to our target; this has been identified as a $z\sim3$ galaxy with prominent emission lines ($\lambda_{\rm obs}\sim2\mu m$) and a faint continuum.
Over the wavelength range 3-3.75$\mu m$ the average continuum flux density of Gz9p3 is greater than that of the neighbouring source by a factor of 5 over the same window. 
The dispersed light from this second galaxy, which is another GLASS target, comes from an open shutter in a separate quadrant of the NIRSpec MSA. The dispersed light is spatially offset and the continuum is sufficiently faint as to not significantly contaminate our target, this can clearly be seen below the Lyman break where the continuum is consistent with zero (i.e. no contaminating emission). 
The weight of the neighbour's continuum emission is lower in the optimal extraction, reducing its influence even more.
However, the strong rest-optical line emission does leave a negative imprint on the continuum trace of our object due to our background subtraction process. These contaminated regions are masked in Figure \ref{fig:spec} and when modeling the spectrum of our object. The close proximity of the dispersed light from other open shutters demonstrates the potential dangers of over filling the NIRSpec MSA.

In the 2D and 1D spectrum of Gz9p3, several emission lines are visible at $\sim4\mu m$ and we present a close up of this region in Figure \ref{fig:spec} with individual panels focusing on each of the detected emission lines. In these panels it can be seen that the $R\sim2700$ resolving power can deblend the \Oii\ doublet into the separate $\lambda_{\rm rest}=3727$ and $3730$\AA\ lines.

Finally, we perform a second extraction of the F100LP/G170H 2D spectrum with the intention of maximizing the signal to noise in the continuum around the observed spectral break ($\lambda_{\rm obs}\in1.2:1.3\mu m$), which we model in  \ref{sec:redshift}. For this purpose we restrict the width of the Gaussian profile ($\sigma=0.1", \sim0.5px$) to reduce the inclusion of pixels that contain only noise.

\subsubsection{Direct Imaging}
Our spectroscopy is complemented by direct \jwst/NIRCam imaging in the F115W, F150W, F200W, F277W, F356W and F444W broadband filters obtained by the DDT program 2756 (PI W. Chen), and reduced and discussed in \cite{Paris23}. We additionally use the HST ACS and WFC3 F606W, F814W, F105W, F125W and F160W broadband filters from \cite{Castellano2016}. All the direct imaging has been PSF-matched to the NIRCam F444W broadband filter, the longest wavelength filter with the coarsest resolution, using the ABELL 2744 region UNCOVER PSF models (epoch: 2022/11/07, PA: 41.2 deg, \cite{Paris23, Bezanson22}).
We report the Kron flux measurements for each broadband filter in Extended Data Table \ref{tab:photometry}, based on the analysis by  \cite{Paris23}. In Figure \ref{fig:spec}, we present the direct imaging of the target in each of the available \jwst\ filters. Here it can be seen that the galaxy exhibits a morphology comprised of a main body plus an extended tail. 

We examine the SED of the full broadband photometry in \ref{sec:SED} and for specific regions within the galaxy using aperture photometry in \ref{sec:SED_regions}.

\begin{table}
\renewcommand{\tablename}{Extended Data Table.}
\renewcommand\thetable{1}
\centering
{%
\begin{tabular}{cc}
Filter & F$_{\nu}$ ($\mu$Jy)\\ \hline \hline
F606W      & $0.01\pm0.02$  \\
F814W       & $0.02\pm0.11$  \\
F105W     & $-0.02\pm0.04$ \\
F125W      & $0.07\pm0.03$  \\
F160W       & $0.31\pm0.05$  \\ \hline 
F115W       & $0.03\pm0.01$   \\
F150W       & $0.32\pm0.01$   \\
F200W       & $0.34\pm0.01$   \\
F277W       & $0.33\pm0.01$  \\
F356W       & $0.32\pm0.01$  \\
F444W      & $0.35\pm0.02$ \\ 
\end{tabular}%
}
\caption{Observed broadband flux density (zero point: zp\_AB=23.9). HST photometry is presented above the horizontal line and \jwst\ photometry is presented below.}
\label{tab:photometry}
\end{table}

\subsection{Gravitational lensing}\label{sec:lensing}
Gz9p3 lies within the ABELL 2744 cluster region, albeit at a relatively large distance from the center of the cluster. Nonetheless this target will experience some degree of magnification. 
In this study we account for the cluster-induced magnification factor at the location and redshift (see Section \ref{sec:redshift}) of Gz9p3 by adopting a magnification $\mu=1.66\pm0.02$ when appropriate, as in  \cite{Castellano22b}. 
The magnification is derived from the latest strong lensing model of A2744 described in \cite{Bergamini23new}. This model exploits 149 multiple images (121 with a spectroscopic confirmation at $1.03 \leq z \leq 9.76$) over an area of 30 arcmin$^2$ and takes advantage of newly identified multiple images based on NIRCam multiband photometry and Very Large Telescope(VLT)/MUSE spectroscopy. We refer to \citep{Bergamini23new} for a detailed description of the modeling of the mass distribution, as well as for the catalogue of the multiple images. The magnification is sufficient to boost the observed magnitude, however is too low to affect the observed morphology of the galaxy, while the object size is magnified by $\sim 30\%\; (\sqrt(\mu)$. 

\subsection{Photometric properties}

This galaxy has a magnification corrected F444W magnitude of \textbf{$25.56\pm0.06$}. Under the adopted cosmology the distance modulus is 49.9, and using a k-correction of $-2.5log_{10}(1+z)$, the object has an absolute magnitude $M_{AB}=-21.77\pm0.06$. This brightness makes it comparable to the rare bright $z\sim9$ galaxies discovered with HST photometry in the large-area BoRG survey \citep[e.g.,][]{Leethochawalit22}.

The rest-frame 1500\AA\ is covered by the NIRCam/F150W broadband filter (with no detected emission lines that contribute to the flux density) and provides a magnification corrected estimate of  ${M_{UV}=-21.67\pm0.04}$ using the same k-correction as above.

These values place this object as one of the intrinsically brightest galaxies spectroscopically confirmed in the early Universe. We plot this object onto the growing sample of known galaxies at high redshift in $M_{UV}$ - redshift space in Figure \ref{fig:Muv_redshift}. At the other luminosity extreme among $z>9$ sources is a highly magnified $z=9.76$ galaxy also from the GLASS-JWST survey \citep{Roberts-Borsani22b} with an intrinsic $M_{UV}=-17.35$ ($\mu\sim13$), demonstrating the dynamical range capabilities of the GLASS program and of \jwst\ in general.

\subsection{Emission line detection and spectroscopic redshift determination} \label{sec:redshift}

The spectrum of this galaxy shows multiple emission lines and a spectral break in the UV, and both features enable the determination of a spectroscopic redshift.

We detect emission from \Oii, \Neiiia, \Neiiib, \He\ and we measure the corresponding line flux without correcting for slit-losses using the latest version of the \texttt{specutils}
packages in \textsc{Python}. 
The stellar continuum in the high-resolution spectroscopy is well modeled by a simple polynomial after masking regions with detected emission line contribution or contamination from the dispersed light of neighboring targets) with a reduced Chi-square = 1.2. The continuum is subtracted before modeling the emission lines. Note that we neglect possible underlying absorption from stellar Balmer absorption lines, which could affect the \He\ line because we use this line uniquely for redshift determination, which is not affected by any reduction in the equivalent width.
We adopt a bootstrap method to represent the flux of each emission line, we model \Neiiia\ with a single Gaussian profile, and the \Oii\ doublet and the \Neiiib+\He\ complex with a pair of Gaussians. The only constraint we place is that within the \Oii\ doublet both emission lines share the same standard deviation ($\sigma$). 
For $n=10,000$ individual computations we vary each spectrum by applying random gaussian noise to each pixel following the associated uncertainty spectrum and fit gaussian profiles to each identified emission line. 
The $n$ fits provide a resultant distribution of measured line fluxes from which we determine the median line flux as the 50$^{th}$ percentile and the associated uncertainty from the 16$^{th}$ and 84$^{th}$ percentiles.
We report the measured line fluxes, which are not corrected for slit losses or reddening, in Table \ref{tab:properties}.

We likewise use the bootstrap Monte Carlo to determine the best-fit redshift. We record $\lambda_{\rm obs}$ for each of the $n$ Gaussian profile fits for the \Neiiia\ emission line and from the [16$^{th}$, 50$^{th}$, 84$^{th}$] percentile we determine the best fit redshift and $1\sigma$ uncertainty to be $z_{spec}=9.3127\pm0.0002$. 
The choice of using \Neiiia\ was made because it avoids the partial-blending in the \Oii\ doublet and the \Neiiib+\He\ complex.

\subsubsection{Spectroscopic redshift from the Lyman break}\label{sec:lyman_break_redshift}

\begin{figure*}
    \renewcommand{\figurename}{Extended Data Figure.}
    \renewcommand\thefigure{2}
    \centering
    \includegraphics[width=\textwidth]{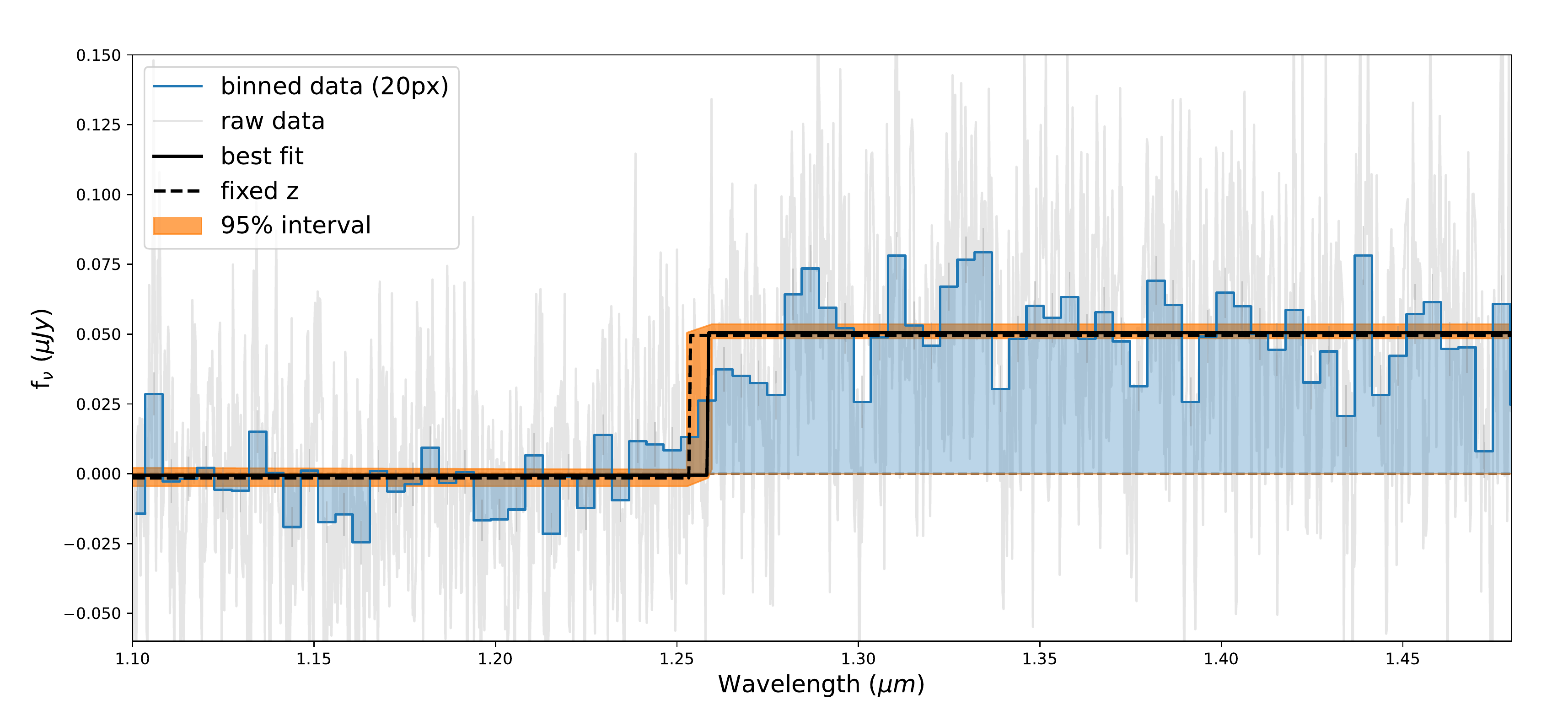}
    \caption{\textbf{NIRSpec constraints on the Lyman break of Gz9p3.} We present a best fit and 95$\%$ confidence interval of $z_{break}=9.35_{-0.05}^{+0.01}$ using the narrow 1D extraction to minimize potential contamination (20px binning presented in blue at the midpoint of each bin, and raw data in gray).
    The orange region covers the $95\%$ confidence interval with the solid black line showing the best fit model. The dashed lines shows the model fit using $z_{spec}=9.313$ derived from the emission lines. The independent best-fit of the Lyman break is consistent with the emission-line spectroscopic redshift solution.}
    \label{fig:lyman_break}
\end{figure*}

We can also measure the redshift from the location of the Lyman break. In the f100lp/g140h spectrum, we attribute the observed spectral break feature (range $\lambda\in[1.2,1.3] \mu m$) to be the Lyman break at $\lambda=1215.67$\AA. To determine the best fit to the spectral break, we use the narrow 1D extracted spectrum (to maximize the S/N) and randomize the data following the uncertainty spectrum by conducting Monte Carlo sampling a total of $n=10,000$ times. We model the Lyman break as a step function and determine a best-fit observed wavelength of $1.258^{+0.001}_{-0.006}\mu m$, 
with the uncertainty assigned by identifying the $95\%$ confidence region when minimizing the $\chi^2$. This best-fit wavelength corresponds to a redshift of $z_{break}=9.35_{-0.05}^{+0.01}$ (in full agreement with the redshift determination from the emission lines). The best fit and uncertainty regions are shown in Extended Data Figure \ref{fig:lyman_break}.  We note that while this break redshift is consistent with the line redshift within the errors, the redshift determined from the break may be systematically slightly higher because of the effect of the Ly$\alpha$ damping wing (see Section \ref{sec:lyman_break}), since we do not consider when we model the spectrum with a step function break.

\subsection{Constraints on other emission lines}

It is also of interest which emission lines fall within our wavelength coverage but are not detected. In the rest-optical we do not detect either HeI lines at $\lambda_{\rm rest}=3890$\AA\ or 4025\AA, placing $1\sigma$ upper limits of 0.50 and 0.39 $\times10^{-19}$ erg s$^{-1} $cm$^{-2}$\AA$^{-1}$ based on the noise properties of the spectrum over a $\lambda_{\rm rest}\pm4$\AA\ window (the extent of the pixel coverage of the detected emission lines). Neither do we detect any of the following commonly-studied rest-UV emission lines: Ly$\alpha$, CIII]($\lambda1907,1909$), MgII($\lambda2804)$, [NeV] ($\lambda3427$) which fall in our coverage (CIV$\lambda1548, 1551$, OIII]$\lambda1661, 1666$, HeII$\lambda1640$, MgII$\lambda2799$ are all out of our coverage). For each of these we place $1\sigma$ upper limits of [2.64, 1.45, 1.07, 0.54] $\times10^{-19}$ erg s$^{-1} $cm$^{-2}$\AA$^{-1}$. We note that many of these are intrinsically faint lines relatively to rest-optical \Oii.

\subsection{Physical interpretation of the spectrum}\label{sec:spec_interp}

The detection of emission lines and the shape of the stellar continuum allows us to infer further galaxy properties of Gz9p3. Although, we note that the location of the MSA open shutter (as shown in Figure \ref{fig:spec}) means we capture the light from the core of the system only.

\subsubsection{Spectrum SED fitting}\label{sec:spec_SED}

\begin{figure*}
    \renewcommand{\figurename}{Extended Data Figure.}
    \renewcommand\thefigure{3}
    \centering
    \includegraphics[width=\textwidth]{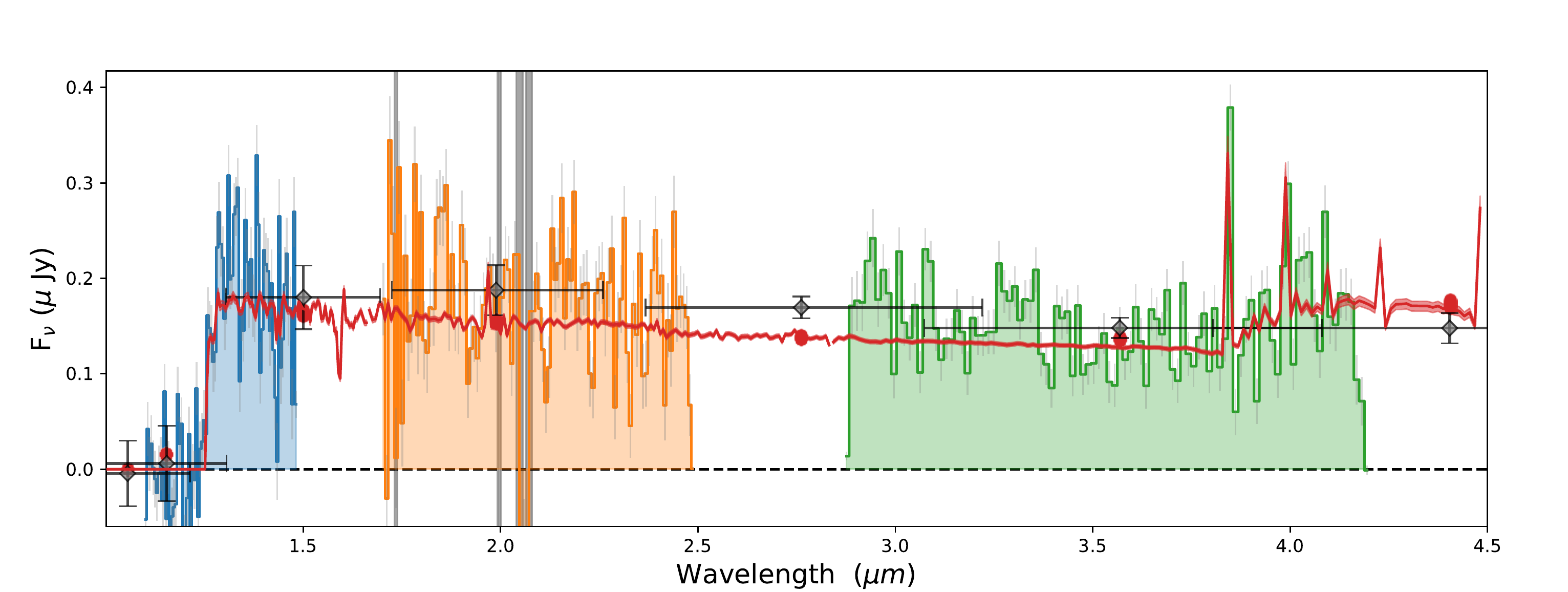}
    \caption{\textbf{Spectral energy distribution fit to the spectrum and photometry of the central region of Gz9p3.} Scaled-spectrum and aperture photometry of the main component of Gz9p3 with best fit BAGPIPES (Spec+Phot) model overlaid (red) on top of the spectrum (both binned at 20px) and the photometric flux densities (where the error bars indicate the filter width and uncertainty on the broadband imaging flux density). Wavelength regions affected by contamination are masked and shaded in gray. Each spectrum has been corrected for slit-losses to match the aperture photometry of the main component (see Extended Data Figure \ref{fig:regions}) in the F150W, F200W and F356W bands respectively. }
    \label{fig:spec_phot}
\end{figure*}

We utilize BAGPIPES \citep{Carnall18} to fit a model SED provided both the spectrum and the photometry to determine the physical properties of the galaxy. For the combined analysis we apply a conversion factor to account for the slit-losses for each of the spectra such that the spectral flux density matches the broadband photometry in the F150W, F200W and F356W filters, respectively for each of the three non overlapping spectral regions covered by the data. Further discussion to validate this correction based on the resolved stellar population is discussed in  \ref{sec:spatial_resolved_SEDs}. The SED fitting procedure is described in \ref{sec:SED}. Extended Data Figure \ref{fig:spec_phot} presents the best fit template overlaid onto the adjusted spectrum and photometry, and the best-fit properties are given in Table \ref{tab:properties}.  We correct our measurements for the lensing magnification; (see ~\ref{sec:lensing}) and we report that the core of the target covered by the shutter has an intrinsic log$_{10}$(M$_*/$M$_\odot$) = $9.15\pm0.04$ and  a 100Myr time-averaged SFR of $9.1\pm 0.6 M_\odot/yr$. The core also shows minimal visual extinction with 
$A_v = 0.14\pm0.04$, equivalent to $E(B-V)=0.57\pm0.20$ for a $R_v=4.05$ \citep{Calzetti00}.
The mass-weighted stellar age of $120\pm20$Myr reflects the presence of an older stellar population and indicates that star formation started as early as $z\gtrsim 15$ to produce the average age observed. 

\subsubsection{Metallicity}\label{sec:metallicity}

The detection of both \Oii\ and \Neiiia\ allows us to study both the number density with the galaxy's ISM, via the ratio of the \Oii\ doublet line fluxes, and the ionization parameter and metallicity from the flux ratio of the \Oii\ and \Neiiia\ lines. We measure a Ne3O2 (defined as \Neiiia/\Oii) line flux ratio of $0.81\pm0.09$. Considering how close these two lines are in wavelength, the measured ratio has only minimal sensitivity to either dust reddening or slitloss correction, which we therefore neglect. 

We determine a metallicity estimate for the galaxy from the measured Ne3O2 line ratio (where we take solar metallicity to be $12 + \log_{10}$(O/H)=8.69, \citep{Asplund09}). The Ne3O2 diagnostic exhibits an anti-correlation with Oxygen abundance \citep{Bian18, Jones15, Shi07, Nagao06, Perez-montero07} due to its relation with the ionization parameter (which is anti-correlated with metallicity, \cite{Jeong20, Levesque14}). Many studies have modeled the relation between the Ne3O2 diagnostic and oxygen abundance in samples of local galaxies. Using 
diagnostic calibrations from literature we determine metallicities of: $7.87\pm0.02$ \citep{Jones15} $7.86\pm0.03$ \citep{Bian18}, $7.38\pm0.07$ \citep{Maiolino_2008}, $7.17\pm0.06$ \citep{Shi07}, with a mean and standard error from these diagnostics of $7.6\pm0.2$. Note that with the data available, we are unable to break the degeneracy in the non-monotonic behavior of the \cite{Nakajima22} relation and obtain two possible values of $7.01\pm0.07$ and $7.88\pm0.07$. Also, our Ne3O2 value appears to lie above the calibrated range of \cite{Sanders21}, placing an upper limit of $<7.4$. All the values quoted here are subject to systematic uncertainties in the diagnostic calibration of the order $0.3dex$, which dominate the random uncertainty.

Our Ne3O2 line therefore supports an oxygen abundance of 10~\% solar ($12+log_{10}($O/H)$=7.6\pm0.2 \pm0.3sys$), in line with recent \jwst\ studies of $z\sim6-10$ galaxies \citep{Sanders23, Mascia23, Cameron23, Nakajima23}.
However, an important caveat is that these metallicity results are based on diagnostics calibrated at low redshifts (SDSS $z<0.1$). While  criteria were used in these studies to select low-redshift EOR analogues, it has not yet been confirmed as to whether these relations still hold true for high-redshift galaxies at $z\gtrsim9$.

\subsubsection{Electron Density}
Following \cite{Sanders16}, we determine an electron density of $n_e = 590^{+570}_{-250} cm^{-3}$ from a measured \Oiib/\Oiia\, flux ratio of $0.94^{+0.14}_{-0.18}$. This has a large uncertainty as $n_e$ is highly sensitive to the flux ratio and the latter measurement has limited signal to noise. \cite{Sanders16} report a median \Oiib/\Oiia\, flux ratio of 1.18 in their $z\sim2.3$ sample ($n_e=225cm^{-3}$). Gz9p3 shows stronger \Oiia\, relative to \Oiib\, than this lower redshift sample, indicating a higher electron density.

\subsubsection{Ionization state} 
The Ne3O2 ratio in this object is significantly higher than has been found in lower redshift samples. The stacked ratio from 66 MOSDEF galaxies of similar mass (log$_{10}$(M$_*/$M$_\odot$):$8.23-9.51$) at redshift $z\sim2$ is Ne3O2 $=0.19$ \citep{Jeong20}, significantly lower than our value. This trend for a high Ne3O2 ratio has also been found in the recent \jwst\ spectra of three $z\sim8$ galaxies ($0.5<$Ne3O2$<2$, \cite{Katz23, Curti23}) in the SMACS early-release observations (\jwst\ program ID: 2736, \cite{Pontoppidan22}),
and is indicative of a higher ionization parameter \citep{Levesque14}. Using the \cite{Witstok21} Ne3O2-U calibration, we determine an ionization parameter of log $U=-2.13\pm0.05$.

Due to a similar dependence on the ionization parameter, Ne3O2 can be used to trace the O32 (\Oiii/\Oii) strong line diagnostic, with the advantage of being less sensitive to reddening \citep{Jones15, Levesque14, Paalvast18, Witstok21}. In a sample of $z\sim0.8$ galaxies, \cite{Jones15} demonstrate that the Ne3O2 diagnostic traces the dust-corrected O32, and can be reasonably approximated as \Neiiia$\approx$\Oiii$/13$, although we note scatter around this relation may hold a dependence on the hardness of the ionizing spectrum \cite{Strom_2017, Zeimann15}. Adopting this approximation would give Gz9p3 a dust-free O32 $=11.2\pm1.3$.

This estimate is larger than the typical range of values found in local samples (O32 $\sim0.3-4$, for star forming galaxies selected from the Sloan Digital Sky Survey \cite{York00} Data-Release 7 \citep{Abazajian09}) 
and falls into the trend that star forming galaxy samples at higher redshift typically exhibit large O32 compared to the local Universe \cite[e.g.,][]{Sanders23, Strom_2017, Runco21}.
Our estimate is comparable to recent \jwst/\NIRSpec\ observations of $z_{spec}\sim8$ sources. 
We find consistency with the range of O32 values (O32$\sim8-20$, \citep{Katz23, Curti23}) found in three SMACS galaxies, the range of O32 values (O32$\sim5-21$, \citep{Mascia23}) found in five GLASS galaxies, the lower O32 limits (O32$>5-8.2$, \citep{Fujimoto23}) placed on six galaxies from the Cosmic Evolution Early Release Science 
survey (CEERS, \jwst\ Program ID: 1345, \cite{Finkelstein22}), and the range of O32 values (O32$\sim6-30$, \citep{Cameron23}) found in five galaxies from the \jwst\ Advanced Deep Extragalactic Survey (JADES, \jwst\ program IDs: 1180, 1181, 1210, 1286 $\&$ 1287, \cite{Eisenstein23}). 

The ISM conditions observed in Gz9p3 provide additional evidence for a higher ionization parameter at high redshift, likely due to the presence of a young stellar population and a harder ionizing spectrum, in line with the trend for an increase in the global specific star formation rate with redshift \citep{Speagle14, Marmol16}.

\subsubsection{AGN constraints}\label{sec:popIII_AGN}
We report no evidence for Gz9p3 being an active galactic Nuclei (AGN). There is no broad-line component required in the model fitting of the detected emission lines, and we place an upper limit of 1.07$\times10^{-19}$ erg s$^{-1}$ cm$^{-2}$\AA$^{-1}$ on the MgII line flux. This corresponds to an equivalent width limit $EW_{\rm rest}<1.9$\AA. Based on a local sample of AGN, this limit would indicate $m_{BH}<10^7~\mathrm{M_{\odot}}$ and a bolometric luminosity of the source far exceeding the Eddington luminosity of the central BH \cite{Dong09}. The inference that an AGN is not the dominant source of light is also fully supported by the extended, spatially resolved nature of Gz9p3.

\subsubsection{Lya absorption}\label{sec:lyman_break}
The shape of the Lyman break can be softened in the presence of a high neutral fraction in the Intergalactic medium (IGM) due to the strength of the Ly$\alpha$ damping wing, allowing the Gunn-Peterson absorption to extend redward of $\lambda_{\rm rest}=1216$\AA\ \citep{Miralda_escude98}. 
To assess whether there is any evidence for Ly$\alpha$ absorption we
estimate the expected continuum level by fitting a $f_\lambda \propto \lambda^\beta$ model to the full stellar continuum in the narrow spectrum (to ensure we avoid any contamination from neighboring spectra) after masking the region of Ly$\alpha$ absorption, regions covering detected emission lines and regions of contamination from emission lines in close-proximity dispersed spectra. Based on the best-fit model we expect a continuum signal at $\lambda_{\rm rest}=1216$\AA\ of $1.01\pm0.03$ $\times10^{-20}$ erg s$^{-1} $cm$^{-2}$\AA$^{-1}$, which agrees with the mean signal over the wavelength range $\lambda_{\rm rest} =1240$\AA$-1300$\AA\ of $<f_\lambda > = 1.02\pm0.05$ $\times10^{-20}$ erg s$^{-1} $cm$^{-2}$\AA$^{-1}$. Over a $\lambda_{\rm rest}=4$\AA\ region redward of Ly$\alpha$ we measure a mean deficit against the expected continuum of $-0.8$ $\times10^{-20}$ erg s$^{-1} $cm$^{-2}$\AA$^{-1}$ (missing 80$\%$ of the expected continuum flux) at a $5.7\sigma$ significance (based on the noise properties of the spectrum). Over a broader $\lambda_{\rm rest}=24$\AA\ region we measure a mean deficit of $-0.4 \times10^{-20}$ erg s$^{-1} $cm$^{-2}$\AA$^{-1}$ at a $6.5\sigma$ significance (40$\%$ of the expect continuum flux).
Our findings are at face value consistent with the expected transmission due to damping wing in a neutral IGM from \cite{Mason20}, with a predicted flux of $\sim0.3\times$continuum and $\sim0.8\times$continuum at 1000 $km\,s^{-1}$ ($\sim4$\AA) and 6000 $km\,s^{-1}$ ($\sim24$\AA).
From this, the flux deficit can be interpreted as Ly$\alpha$ absorption and thus as evidence that this galaxy does not sit within a large ionized bubble. Similar interpretations have been drawn for 4 galaxies at $10<z_{spec}<14$ which also show a softening of their Lyman break \citep{Curtis-lake22}. However, this physical interpretation is challenging to reconcile with the evolved stellar populations and large stellar mass estimates, which imply the reionization process of the IGM should be well underway. In this respect, we highlight that Ly$\alpha$ has been observed in emission in GN-z11, a younger, less massive galaxy at higher redshift \cite{Bunker23}. An alternative interpretation of the Ly$\alpha$ damping is explored in the context of ISM and CGM absorption in \ref{sec:absorption_lines}. 

\subsubsection{UV absorption features}\label{sec:absorption_lines} 
\begin{figure*}
    \renewcommand{\figurename}{Extended Data Figure.}
    \renewcommand\thefigure{4}
    \centering
    \includegraphics[width=\textwidth]{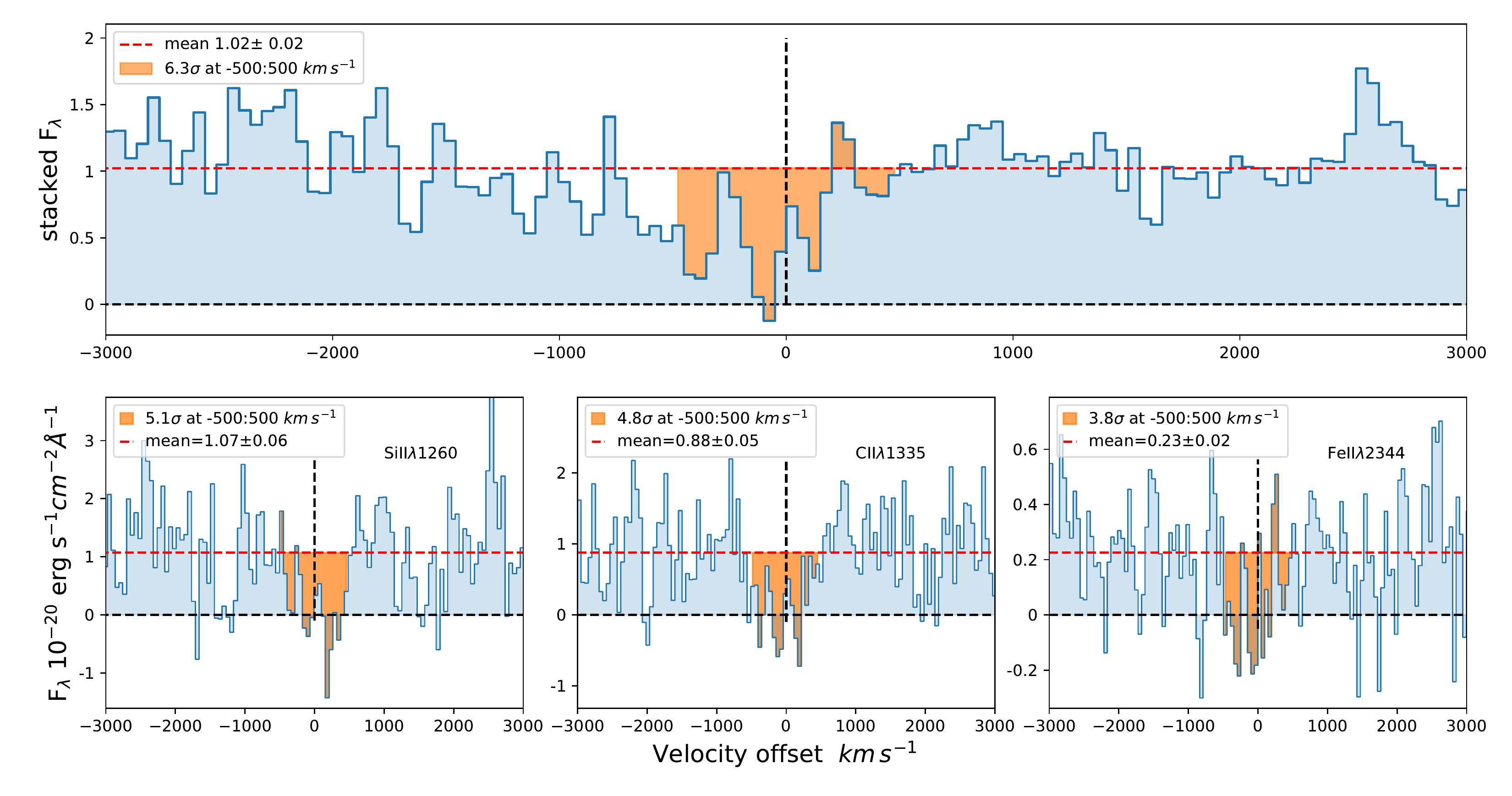}
    \caption{\textbf{UV absorption features in the high-resolution spectroscopy of Gz9p3.} Top panel: Stack of region $\pm3000km\,s^{-1}$ centered on common UV-absorption lines (SiII$\lambda1260,1304$, OI$\lambda1302$, CII$\lambda1335$, FeII$\lambda2344,2374,2382$). The orange filled region highlights the $-500:500km\,s^{-1}$ window which shows a series of absorption features exhibiting a 40$\%$ reduction flux compared to the mean stellar continuum (red line) at a $6.3\sigma$ significance. Bottom panels: the individual SiII$\lambda1260$, CII$\lambda1335$ and FeII$\lambda2344$ absorption features shown over the same velocity window.}
    \label{fig:uv_absorption}
\end{figure*}

We apply the same method used for Ly-$\alpha$ absorption to measure the significance of a deficit between the expected stellar continuum and the measured continuum for several common UV absorption lines which fall within our spectral coverage (SiII$\lambda1260,1304$, OI$\lambda1302$, CII$\lambda1335$, FeII$\lambda2344,2374,2382$).
We examine the region of $\pm5000km\,s^{-1}$ centered on each line and measure the average continuum level outside the central $\pm500km\,s^{-1}$. We measure the significance of any absorption within this central region against the continuum level and find greater than $>3\sigma$ detections in SiII$\lambda1260$, CII$\lambda1335$ and FeII$\lambda2344$, with $EW_{\rm rest} = (-3.6\pm0.7)$\AA, $(-3.7\pm0.8)$\AA\ and $(-6.2\pm1.6)$\AA, respectively. The remaining lines show $<2\sigma$ significance. 
To examine potential UV-absorption further we stack the $\pm5000 km\,s^{-1}$ window for these 7 lines, scaling the measured continuum level to unity and using a $1/\sigma^2$ weighting. In Extended Data Figure \ref{fig:uv_absorption} we present the stacked spectrum plus the individual spectral regions for the detected absorption lines. 
The stacked spectrum shows a double absorption feature centered at $\pm100 km\,s^{-1}$. The deficit of the combine absorption feature over a [-500:500]$km\,s^{-1}$ window is a 40$\%$ reduction flux compared to the stellar continuum at a $6.3\sigma$ significance (57$\%$ reduction at $7.5\sigma$ over a $[-500:200]km\,s^{-1}$ window). 

The equivalent width of SiII and CII absorption lines is closely correlated with the Ly$\alpha$ equivalent width as these ions arise from HI gas, which is also responsible for the damping of Ly$\alpha$ \cite{Jones2012, Leethochawalit2016, Pahl2020, Du2018, Shapley03}. From \cite{Pahl2020}, we estimate that rest-frame equivalent width of $\lesssim -2$ \AA\ for low-ionization interstellar metal lines is associated with weakened Ly$\alpha$ emission and moderate absorption damping wings. Therefore, our measurement of SiII and CII absorption with $EW\sim (-3.6 \pm 0.8)$\AA~ can qualitatively explain the damped Ly$\alpha$ feature present in the spectrum redward of the Lyman break without necessarily requiring Gz9p3 to live within a partially neutral IGM region. While our data are inconclusive to determine the IGM ionization, we note that the combination of strong absorption for low-ionization ISM metal lines and weak CIII] emission (EW$_{\rm rest} < 1.0$\AA) suggest that the galaxy is unlikely to be a Lyman continuum leaker in its current state \cite{Shapley03,Lopez2022}.

In both the stack and the individual lines the overall shape of the deficit shows hints of a multi-component structure. The absorption is centered at a $100km\, s^{-1}$ offset to the expected location, either to one side or both. Such multi-component absorption may indicate the presence of turbulent gas within the system, but the signal to noise of our data is not sufficient to draw robust conclusions on this potential interpretation. 

\subsection{Measurements of integrated physical parameters} \label{sec:SED}
Given that the NIRSpec slit does not cover the full extend of the object due to its orientation (see Figure ~\ref{fig:spec}), we additionally use integrated-light photometry to derive an estimate for the stellar mass and the 100Myr time-averaged star formation rate (SFR) for the whole galaxy by fitting the spectral energy distribution using two alternative SED-fitting software tools; BAGPIPES \citep{Carnall18} and ZPHOT \citep{Fontana00}.

The application of each tool for the GLASS photometric dataset is described in \cite{Leethochawalit23} for BAGPIPES and \cite{Santini2022} for ZPHOT, and we refer the reader to these articles for a detailed discussion.  Briefly, for BAGPIPES, we model the photometry using a lognormal SFH, a Kroupa \cite{Kroupa02} initial mass function (IMF), a Calzetti \cite{Calzetti00} dust attenuation law and a BPASS (v2.2.1 \cite{EldridgeStanway2009}) stellar population model that includes binary populations that is reprocessed with photoionization code CLOUDY (c17.03 \cite{Ferland2017}) generated in \cite[][private communication with Adam Carnall]{Hsiao2022}. Consistent stellar mass and SFR estimates within $1\sigma$ are obtained using a delayed exponential SFH or when using the 2016 version of the BC03 \cite{Bruzual03} stellar population models.
For ZPHOT, the important changes are the use of a delayed exponentially declining Star Formation Histories, the 2016 version of the BC03 \cite{Bruzual03} stellar population models and a Chabrier \cite{Chabrier_03} IMF. 

In Section \ref{sec:metallicity} we place constraints on the metallicity of Gz9p3 and as part of our SED fitting we apply a Z=[0.01-0.33]Z$_\odot$ flat prior to match the measured range of metallicities determined from our Ne3O2 analysis and their associated systematic uncertainty. We note that we obtain consistent measurements (within $1\sigma$) for our key parameters (Mass, SFR, Age) when we place no prior on the metallicity.

We run both tools twice, first allowing the redshift to vary and then  fixing the redshift at $z_{spec}=9.313$ (see \ref{sec:redshift}). All variations of the photometric analysis return consistent estimates of the stellar mass and star formation rate, and we find the photometric redshift estimates for both tools are reasonably consistent with the spectroscopic value and place the redshift firmly at $z>9$ ($z_{phot}$: $10.2$ and $9.7\pm0.3$ for ZPHOT and BAGPIPES respectively).
By fixing the redshift at the solution derived from emission line analysis, BAGPIPES and ZPHOT estimate an observed log$_{10}$(M$_*/$M$_\odot$) of $9.43^{+0.11}_{-0.15}$ and $9.51^{+0.20}_{-0.14}$ and SFRs ($M_\odot/yr$) of $31^{+8}_{-10}$ and $57^{+9}_{-21}$. The consistency between the tools is discussed in \cite{Santini2022}, who find good agreement in all but the lowest mass galaxies. We present the best-fit SED using the BAGPIPES fixed-redshift model in Extended Data Figure \ref{fig:regions}.

We correct our measurements for the lensing magnification ($\mu=1.658^{+0.016}_{-0.015}$; see ~\ref{sec:lensing}) and we report that our target has an intrinsic log$_{10}$(M$_*/$M$_\odot$) $=9.2^{+0.1}_{-0.2}$ (or $9.3^{+0.2}_{-0.1}$) and SFR $=19^{+5}_{-6} M_\odot/yr$ (or $34^{+5}_{-13} M_\odot/yr$) for BAGPIPES (or ZPHOT). These measurements are consistent with those obtained by \cite{Castellano22b} using ZPHOT, with \cite{Charlot00} templates and a delayed-star forming history (SFR=$40^{+70}_{-10} M_\odot/yr$, log$_{10}$(M$_*/$M$_\odot$)=$9.2\pm0.3$).

\subsubsection{Mass metallicity relation}
This galaxy exhibits a large stellar mass and low oxygen abundance. In Figure \ref{fig:mass_metal} we present the location of Gz9p3 relative to measured mass-metallicity relations. The measured metallicity range for Gz9p3 [$12+\log_{10}($O/H$) = 7.2-7.8$] places it offset below local $z=0$ \citep{Andrews13, Maiolino_2008} and moderate redshift $z=2-4$ mass-metallicity relations \citep{Sanders21, Maiolino_2008} (which estimate $12+\log_{10}($O/H$)$ in the range $\sim8.5-8.8$ and $\sim7.7-8.3$ at $\log_{10}$(M$_*/$M$_\odot$)$=9.2$, respectively). In a recent census of \jwst\ galaxies at $z\sim4-10$, \cite{Nakajima23} report little evolution in the mass-metallicity relation within their metallicity uncertainties ($\sim0.3dex$) and we find  Gz9p3 remains offset below the best-fit mass-metallicity relation at these higher redshifts. In fact, the relation from \cite{Nakajima23} provides $12+\log_{10}($O/H$) = 8.01$ at  $\log_{10}$(M$_*/$M$_\odot$)$=9.2$. 

\subsection{Resolved stellar populations} \label{sec:spatial_resolved_SEDs}

In Figure \ref{fig:spec}, Gz9p3 is clearly isolated from close neighbors and shows an elongated spatially-resolved structure with relatively high S/N in individual pixels. 
The direct images have been PSF-matched to the F444W image and while the PSF-matching means the information encoding within a pixel will be blended with its neighbors, our target is spatially large enough to provide regions far enough apart to be resolved (FWHM of F444W is $\sim4.5$ native pixels, or $\sim0.14"$). This offers the opportunity to carry out a resolved stellar population analysis.
Here, we follow the method of \cite{Gimenez-Arteaga22} and perform a pixel-by-pixel fitting of the SED. By fitting the physical parameters for each pixel we can create a 2D map of the distribution of galaxy properties (Stellar Mass, SFR, Stellar Age).

To ensure we have sufficient sensitivity to perform pixel-by-pixel SED model fitting, we first bin the direct image using a (2$\times$2) kernel to improve the signal to noise (SNR) and then enforce a $SNR>2$ threshold in the F150W and F200W \NIRCam\ broadband filters (which have the highest noise levels). After this binning, there are $\sim60$ pixel that meet this threshold, which sample the full extent of the object with a (binned) pixel scale of 0.062"/px, which at $z=9.313$ scales to 0.27kpc/px.

For each pixel we fit the 11-band HST+\jwst\ photometry using BAGPIPES (as discussed in \ref{sec:SED}) and we fix the redshift to $z_{spec}=9.313$. Our $SNR>2$ threshold choice is sufficient to obtain robust fits, with the log of the Bayesian evidence for all pixels $>2$.

From BAGPIPES we obtain estimates for the star formation rate and stellar mass, which we convert to surface density properties using the 0.27 kpc/pixel scale, as well as an estimate for the mass-weighted stellar age. 
We additionally measure the UV slope $\beta$ (where $f_\lambda\propto\lambda^{\beta}$) using the F150W-F200W colour, taken from the BAGPIPES best-fit SED magnitudes.
We determine $\beta = C\times (m_{1} - m_{2}) - 2$ (e.g., \cite{Dunlop13}), where $C^{-1}=2.5\cdot log_{10}(\frac{\lambda_2}{\lambda_1})$. For F150W-F200W, the effective wavelength $\lambda_{eff}=15007$\AA\ and $19886$\AA\ trace the rest-frame UV for our galaxy's redshift ($\lambda_{\rm rest}=1455$\AA$-1930$\AA), and set $C=3.27$. Here we adopt the \lq pivot\rq\ wavelength as the effective value to account for the transmission profile and detector sensitivity as a function of wavelength for each filter.
In Figure \ref{fig:maps} we present the 2D distribution of these physical properties, where for each pixel we display the 50$^{th}$ percentile of the posterior and in the supporting panels we show the uncertainty of the estimates based on the 16$^{th}$ and 84$^{th}$ percentiles.

Within the 2D distributions, the presence of multiple components is apparent; the main body of the object, a tail and a compact region at the end of the tail (highlighted in the F150W-F444W panel of Figure \ref{fig:maps} by the $10\sigma$ F150W contour). 
The SFR surface density peaks in the center of the main body, with minimal star formation found in the tail. 
The mass-weighted age for the main body and tail-clump is estimated to be $<50Myr$, with the tail showing older ages albeit with a larger uncertainty. 
Across the whole 2D distribution of the galaxy the estimates for the visual extinction are below $A_v=0.7$, with a median of only $A_v=0.2$. It is clear that the SED modeling prefers solutions with little to no dust extinction.
Like the SFRD, the stellar mass surface density peaks in the center of the main body with a SMD of $\sim10^6M_\odot/kpc^{-2}$ maintained throughout the tail. 
The $\beta$ slope estimates range from $-1.5$ to $-2.5$ and trace the stellar age 2D distribution, with the steepest slopes (associated with the youngest stellar populations) correlating with the stellar age. 

\subsubsection{Properties of Integrated regions}\label{sec:SED_regions}

\begin{figure*}
    \renewcommand{\figurename}{Extended Data Figure.}
    \renewcommand\thefigure{5}
    \centering
    \includegraphics[width=\textwidth]{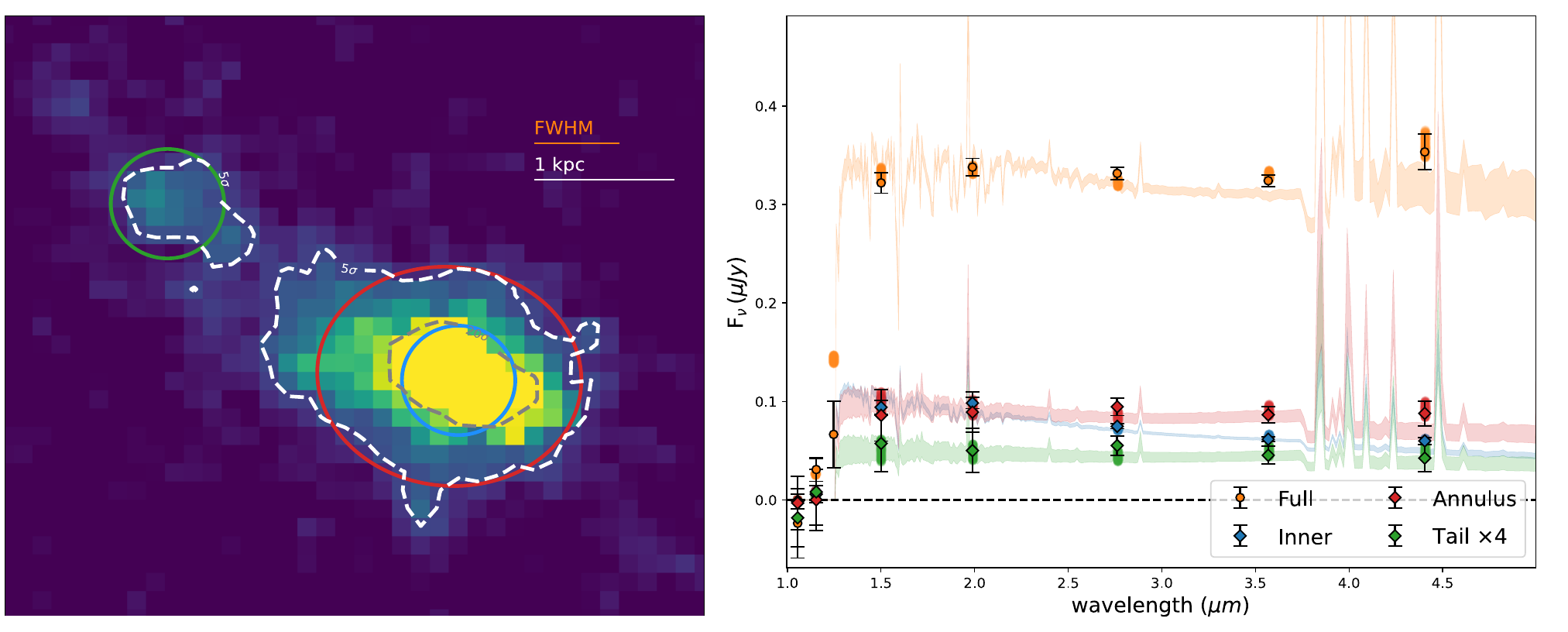}
    \caption{\textbf{Spectral energy distribution for Gz9p3 from integrated light in key regions.} Left Panel: Overlaid onto the F277W direct imaging for the galaxy, white dashed lines show the $5\sigma$ and $20\sigma$ F277W contours. Three apertures are placed to approximately trace these contours. These include an aperture (green) over the tail tracing the $5\sigma$ contour of a clump and an inner and an outer aperture over the main component tracing the $20\sigma$ and $5\sigma$ contours (red, blue). 
    Right Panel: Photometry and BAGPIPES SED fit ($16^{th}-84^{th}$ percentile shown by shaded region) for the full galaxy (orange, see Extended Data Table \ref{tab:photometry}) and key regions from the left panel 
    (blue $=$ Inner, red $=$ the annulus region created between the inner and outer apertures on the main source, and green $=$ Tail, which we scale by $\times4$ to aid the viewer). Error bars derive from uncertainty on the broadband imaging flux density and the SED fitting. }
    \label{fig:regions}
\end{figure*}

In addition to presenting the pixel-by-pixel 2D distributions of the physical parameters, we investigate the integrated properties of three distinct regions (with greater spatial separation than the FWHM). Extended Data Figure \ref{fig:regions} presents the placement of three apertures which follow the 5 and $20\sigma$ contours of the F277W direct image. One aperture ($r=3px$) is placed over clump in the tail, which traces the region of low stellar ages and blue $\beta$ slope seen in Figure \ref{fig:maps}. Two apertures are placed over the main component of the source, a compact aperture ($r=3px$) following the $20\sigma$ contour and a larger ellipse ($6px/7px$ semi-minor/major axis) following the $5\sigma$ contour, the placement of these two creates an annulus. The inner aperture traces the compact peak of the SFRD and SMD distributions while the annulus covers the surrounding lower SFRD and SMD region.

To compare different regions of the target, we use the
\texttt{photutils}
package in \textsc{Python} to measure the aperture photometry.
We perform BAGPIPES SED modeling of the aperture photometry from each region following the method in \ref{sec:SED}, obtaining tighter constraints on the physical parameters than in the pixel-by-pixel fitting. The photometry and best-fitting SED models are shown in the right panel of Extended Data Figure \ref{fig:regions}.

The Inner region and the surrounding annulus have comparable F150W and F200W flux densities, however diverge at longer wavelengths, with the inner region showing a steeper UV-slope ($\beta=-2.48^{+0.05}_{-0.03}$ and $-2.1\pm0.2$, respectively) and a fainter rest-optical stellar continuum. The clump in the tail shows similar properties to the annulus region, with a $\beta=-2.2^{+0.4}_{-0.2}$ UV slope.
All three regions exhibit consistent sSFR ($\sim10Gyr^{-1}$) and Stellar ages ($\sim8$Myr, with 16$^{th}/84^{th}$ percentiles of $\sim3$ and $\sim20$Myr), with the distinguishing property being the visual extinction which is significantly lower in the Inner region ($A_v = 0.03^{+0.04}_{-0.02}$) than in the annulus and tail ($A_v = 0.3\pm0.2$ and $0.3^{+0.3}_{-0.2}$, respectively). 
In each region the extinction is minimal and for the Inner region it appears negligible, possibly suggesting inflow of pristine gas to the core due to a merger event (see Section \ref{sec:merging}). 

The inferred properties of these regions from the photometry alone are presented in Extended Data Table \ref{tab:region_results} and all show young stellar ages ($<10Myr$), whereas the fit to the spectrum and photometry of the main component (the combined inner+annulus region) in \ref{sec:spec_interp} estimated a higher age ($130\pm20$Myr). These differences highlight how the light from the young stellar populations dominates the photometry in the rest-UV and masks the older population, which can be seen in the spectra based on the emerging Balmer break.
Likewise, the stellar mass estimate for the main component using the spectrum+photometry  (log$_{10}$(M$_*/$M$_\odot$)=$9.15\pm0.04$) is higher due to the inclusion of an older stellar population. 
From the photometry-only SED fitting, the stellar mass of the main component accounts for about one fifth of the full-photometry estimate, but for half of the $M_{UV}$ flux density. Correspondingly, the outskirts of the main component (outside the aperture) and the extended tail exhibit a high stellar mass-to-UV-light ratio. This follows the pattern of SMD, stellar age and SFRD properties in the 2D distributions of Figure \ref{fig:maps}. We also note that the very young stellar population ages ($<10Myr$) preferred in the fits of the individual regions have lower stellar mass-to-UV-light ratios, with the fit outcome dominated by the rest-frame UV light. The discrepancy in the star formation rates and stellar masses between SED fitting from broadband photometry and SED fitting from the spectrum highlights the importance of spectroscopy to study galaxies during the epoch of reionization. 

\begin{table}
\renewcommand{\tablename}{Extended Data Table.}
\renewcommand\thetable{2}
\centering
\resizebox{\columnwidth}{!}{%
\begin{tabular}{ccccc}
Property            &  Main (phot-only) &  Inner &  Tail &  Annulus \\ \hline
Stellar Mass [log$_{10}($M$_*/$M$_\odot)$]       &
$8.3^{+0.3}_{-0.2}$
&
$7.9\pm0.1$
&
$7.3^{+0.3}_{-0.2}$
&
$8.2^{+0.3}_{-0.2}$    \\
Star formation rate [M$_\odot\,yr^{-1}$] &
$2.2^{+2.5}_{-0.7}$
&
$0.9^{+0.4}_{-0.3}$
&
$0.2^{+0.3}_{-0.1}$
&
$1.5^{+1.6}_{-0.6}$      \\
Stellar Age [Myr]         &
$7^{+15}_{-3}$
&
$8^{+7}_{-4}$
&
$8^{+20}_{-4}$
&
$7^{+20}_{-4}$      \\
$A_v$                  &
$0.1\pm0.1$
&
$0.03^{+0.04}_{-0.02}$
&
$0.3^{+0.3}_{-0.2}$
&
$0.3\pm0.2$      \\
$\beta$ Slope          &
$-2.3\pm0.1$
&
$-2.48^{+0.05}_{-0.03}$
&
$-2.2^{+0.4}_{-0.2}$
&
$-2.1\pm0.2$      \\
Muv  [AB\_mag] &
$-21.1\pm0.1$
&
$-20.34^{+0.04}_{-0.03}$
&
$-18.2^{+0.3}_{-0.2}$
&
$-20.4^{+0.2}_{-0.1}$
\end{tabular}%
}
\caption{Galaxy properties for regions within Gz9p3, fit using aperture photometry shown in Extended Data Figure \ref{fig:regions}, corrected for magnification when appropriate. The spectrum+photometry SED fitting results for the main component is shown in Table \ref{tab:properties}.}
\label{tab:region_results}
\end{table}

\subsection{Morphological properties and evidence for a merging system}\label{sec:merging}

The non-negligible mass in the tail of Gz9p3 suggests the extended structure is a result of a merger rather than being due to gas stripping, which would show very low masses in a tail \citep{Vulcani21}. This is further supported by the presence of a double nucleus (reminiscent of a merger remnant) in the F150W-F444W panel (and in Figure \ref{fig:spec}) and by star formation concentrated to the core, which may indicate a nuclear starburst triggered by gas infall during a merger \citep{Springel00, Barnes02, Naab06}. On the other hand, the pixel-by-pixel SED modeling returns a relatively smooth spatial distribution of properties, which could be interpreted as evidence that the system is observed a few dynamical times post-merger and has therefore largely completed its relaxation process. 

Spectroscopic coverage over the tail, which is unavailable due to the placement of the slit, would likely allow stellar and gas kinematic measurements to further distinguish between a merger history and any gas stripping from either ram pressure stripping within the $z\sim10$ over-density reported by \cite{Castellano22b} or due to cosmic web stripping \cite{Benitez-Llambay13}.

Lacking such spectroscopic data, to examine further whether this system is a merger, we utilize a tailored data reduction of the NIRCam direct imaging at short wavelengths, PSF matching to F200W (the longest wavelength filter with the coarsest resolution we consider in this Section, instead of F444W). This allows greater spatial resolution in the short wavelength F150W and F200W filters, which are drizzled to a 20mas/px resolution. We weight-combine the F150W and F200W direct images to improve the signal to noise ratio before investigating the morphology. The combined image is presented in Figure \ref{fig:clumpy} and visually shows two cores within the center of the galaxy. We fit both core components (Left and Right as viewed) with Sersic profiles and measure consistent profiles with $Re_{major}= 0.032",  0.033"$(corresponding to $0.14kpc$), $n_{sersic}=0.44, 0.55$ and axis ratios of 0.81, 0.76. The central region (excluding the tail) shows a greater axis ratio of 0.44, with a  $Re_{major}=0.15"$ (0.65kpc) for a $n_{sersic}=0.50$ profile. The best-fit F150W+F200W profile, which traces the rest-UV and hence active star forming regions, shows a smaller size and shallower gradient than the F444W best-fit profile to the central region ($Re_{major}=0.26", n_{sersic}=3.54,$ axis ratio$=0.26$) which traces the rest-optical and the older stellar population.

\subsubsection{Morphological indexes}
Several quantifiable indicators exist for describing the morphology of a galaxy, with thresholds to indicate whether a system is likely to have gone, or be going through, a merger: Gini - M$_{20}$ and asymmetry (A). 
These morphological parameters are defined in \cite{Lotz04, Lotz08} and \cite{Abraham03}, respectively. \cite{Lotz08} define mergers as galaxies with Gini $> 0.33$ $- 0.14 \times$ M$_{20}$, while \cite{Conselice03} adopt the condition $A \geq 0.35$ as a merger criterion. 
These parameters were already adopted to study the morphology of galaxies at the epoch reionization with \jwst\ observations \citep{Treu23}. We thus follow the same measurement procedure on the combined F150W+F200W galaxy cutout shown in Figure \ref{fig:clumpy}. 
We determine each morphological parameter following the literature definitions with our in-house \texttt{JWSTmorph} package.
We obtain Gini $=0.61$, M$_{20} = -1.29$, and $A=0.35$, which robustly classify our system as a merger.

\subsubsection{Clumpiness}
The system appears also very clumpy in Figure \ref{fig:clumpy}, with two bright and distinct clumps clearly detected in the central region. These might represent the two merger components. To be more quantitative, we calculate the clumpiness parameter $c$, which is defined as the fraction of light of the galaxy residing in clumps. This traces the relative importance of small scale structures inside a galaxy, and can shed light on star-formation episodes outside of the nucleus or material (gas+stars) that has been stripped away from the galaxy by tidal phenomena. At intermediate redshift, the presence of bright clumps is tightly associated to merger events \citep{Calabro19}. 
To compute the clumpiness, we follow a similar approach to \cite{Calabro19}. In brief, we smooth the original image (F150W + F200W combined) using a gaussian filter with a size of $0.2''$, corresponding to $\sim1$ kpc at the target redshift. 

We then subtract the smoothed image from the original image, and select the pixels with a flux at least $2\sigma$ above the background, in order to reduce the noise contamination. We also impose $0$ for all the pixels with a negative value in the residual \citep{Conselice03}. Finally, the clumpiness is calculated as the ratio between the flux in those pixels and the total flux of the galaxy. 
We obtain a final value of $c = 0.56$, and the clump-map is shown in the right panel of Figure \ref{fig:clumpy}. 

In the clump-map we can see that, while most of the flux is coming from the two central nuclei, there are $4$ clumps detected in the left tail of the system. If we remove the nuclei, we still find that $6.3 \%$ of the total flux (which roughly traces the SFR as we probe the rest-frame UV) resides in these external structures, which is completely different from the compact, nucleated morphology of typical star-forming galaxies seen at these redshifts \citep{Treu23}. We can also notice the elongated shape of the main clump in the center, which contributes to increase the asymmetry of the system.
All this evidence suggests that a merger is ongoing and that the elongated, clumpy structure might be the result of the tidal forces generated during the interaction, as observed in lower redshift mergers.

\subsection{Comparison to theoretical modeling}\label{sec:theory}

The properties of Gz9p3 can be compared to basic theoretical modeling of galaxy formation in $\Lambda$CDM as a first step to assess the ability of current model to reproduce the observations. To estimate the number density and relation between stellar mass and $M_{UV}$, we consider a set of models where DM halos are populated with stellar populations based on simple analytical recipes \citep{Mason2023,Mason2015,Ren2019}. In summary, the modeling relies on three elements (1) the stellar-to-halo mass ratio is mass dependent but redshift independent; (2) the age of the stellar population is proportional to the DM halo assembly time and its scatter derives from the probability distribution of the halo assembly time based on extended Press Schechter theory; (3) the star formation efficiency is calibrated via abundance matching at a single reference redshift where a robust determination of the UV luminosity function is available. Using the results from \cite{Mason2023} we can see from their Fig.~3 that Gz9p3 lies very close to the modeling predictions for the stellar mass versus $M_{UV}$ relationship at the median age of $\sim 80$ Myr, indicating it should be a typical object for its mass and luminosity (and dust content). Based on that model, the inferred DM halo mass of the system is $M_{DM}\sim 10^{11.4}~\mathrm{M_{\odot}}$ and the inferred number density is $\sim 10^{-6}~Mpc^{-3}$ (comoving). It is challenging to estimate precisely the parent survey volume to associate to this object (given it was included as a NIRSpec target based on pre-existing HST photometry), however bounding within a factor of two is possible. The lower limit is to consider the NIRSpec FoV of $\sim 3\times 3~arcmin^2$  and $\Delta z \sim 1$, while the upper limit is the area of the HST photometry that includes WFC3/IR data (needed given the redshift of the source), which is approximately $13~arcmin^2$ \citep{Treu23}. This gives a selection a volume of $\sim (2-3)\times 10^4~Mpc^{-3}$ (comoving) (see e.g. the cosmic variance calculator from \cite{Trenti08}), and it implies only a few percent probability of having such a massive and bright galaxy in our observations. Given that the relationships between stellar mass,  UV luminosity, and dust are in agreement with theory, the natural degree of freedom to change is an increased star formation efficiency relative to the halo mass as the redshift increases. This has been proposed in similar modeling, e.g. by \cite{Behroozi2015}, and an increased star formation efficiency could also explain the higher than expected density of photometric candidates at $z\gtrsim 10$ from NIRCam observations (e.g. \cite{Castellano22b}). Another possibility is that the source is hosted by a dark matter halo mass that is less massive than the average halo to stellar mass relationship implies, and thus we are observing an outlier with higher than average star formation efficiency and stellar mass. Based on previous work, it is possible that the efficiency in a single object is up to one dex above mean (i.e. a factor $10\times$) (e.g. \citep{Ren2019,Mason2023}).

Theoretical modeling is also useful to estimate the probability of observing a disturbed morphology deriving from a galaxy merger. The merger rate of DM halos per unit redshift is nearly universal across primary halo mass and redshift and only depends on the mass ratio \citep{Fakhouri2008}. Estimating a mass ratio between $1/10$ and $1/5$ from the photometry of Gz9p3, \cite{Fakhouri2008} gives $dN_{merge}/dz\sim 1$, which corresponds  to one merger expected every $\sim 70$ Myr at $z\sim 9.3$. Assuming further that the disturbed morphology remains observable over a timescale of a few dynamical times for the system. The latter can be estimated to be $t_{dyn}\sim 5$ Myr considering a characteristic velocity dispersion of $\sim 200$ km/s (from virial equilibrium) and a characteristic radius of $3$ kpc (from imaging). Therefore, we conclude that catching galaxies in the act of merging at this epoch is relatively likely (at the level of $\sim 10-20$\% probability). 

\subsection{Comparison to cosmological hydrodynamical simulations} \label{sec:simulations}

To further assess whether this galaxy is in tension with predictions from the $\Lambda$CDM model, we searched for analogous galaxies with masses $10^9 \leq M_* / M_\odot \leq 10^{10}$ in the IllustrisTNG simulation suite \citep{TNG1, TNG2, TNG3, TNG4, TNG5}. We found 13 analogous galaxies in Snapshot 5 of TNG300-1 at a redshift of $z=9.39$, with a box length of 302.6 cMpc, 
and an average mass resolution of $1.1 \times 10^7 M_\odot$ for baryonic particles. No analogous galaxies were found in TNG100, the second largest simulation of the suite with a box length of 110.7 cMpc, indicating that such massive galaxies are rare objects in the early Universe.

While 13 simulated galaxies does not provide a statistical sample, we do find that these simulated sources do share similarities to Gz9p3. Each of the 13 galaxies shares a similar SFR, with the mean SFR $= 30\pm9 M_\odot yr^{-1}$. 
The merger tree history of the 13 systems indicates that two had undergone a major merger (Mass ratio $>$ 0.2, 15$\%$) and four had undergone a minor merger (Mass ratio $<$ 0.2, 31$\%$) within the cosmic time interval between $z=10$ and $z=9.3$ (Snapshots 4 and 5 of the simulation). Such a rate of major mergers falls in line with our prediction from theoretical modeling. Unfortunately, the spatial and mass resolution of the simulation is not sufficient to obtain a detailed morphology for a quantitative comparison. However, qualitatively, two simulated systems show a double-core structure (two distinct mass peaks) which reflects the structure of Gz9p3. Therefore, while current state of the art modeling and simulations suggest that the probability of finding an object such as Gz9p3 in the cosmological volume probed by this survey is low (approximately 6\% considering 13 objects in the TNG300 and assuming a survey area of GLASS-ERS+DDT+UNCOVER of $\sim 60$ arcmin$^2$), setting aside number density, galaxies with properties similar to Gz9p3 are expected to exist in the early Universe. This may suggest that star formation efficiency during the epoch of reionization may need to be revised upwards to improve the data-model comparison. 

\section*{Declarations}

\begin{itemize}
\item Data availability: All data used in this paper are publicly available through the Mikulski Archive for Space Telescopes (MAST) server with the relevant program IDs (ERS-1324 for the NIRSpec spectroscopy and DDT-2756 for the NIRCam imaging). The reduced NIRCam imaging utilised in this work from the GLASS collaboration \citep{Paris23} is available at \url{https://doi.org/10.17909/kw3c-n857}. All other data generated throughout the analysis are available from the corresponding author on request. 
\item Code availability: Our analysis makes use of several publicly available codes. 
The NIRSpec data were reduced using the \texttt{msaexp} code, which can be found here: \url{https://github.com/gbrammer/msaexp}. 
The data reduction of the NIRCam images were performed with the official STScI JWST pipeline, which can be found here: \url{https://github.com/spacetelescope/jwst}.
The SED fitting analyses were performed with BAGPIPES, the latest version of which (including the templates used here) is available at \url{https://bagpipes.readthedocs.io/en/latest/}. 
We modelled the observed spectral emission lines using the \texttt{specutils} packages within \textsc{Python}, which can be found at \url{https://specutils.readthedocs.io/en/stable/}.
We performed aperture photometry on the direct imaging using the \texttt{photutils} packages within \textsc{Python}, which can be found at \url{https://photutils.readthedocs.io/en/stable}.
Galaxy morphological parameter were measured using the GLASS in-house \texttt{JWSTmorph} package, which is publicly available at \url{https://github.com/Anthony96/JWSTmorph.git}.
All other code generated throughout the analysis are available from the corresponding author on request. 
\item Acknowledgements: This work is based on observations made with the NASA/ESA/CSA James Webb Space Telescope. The data were obtained from the Mikulski Archive for Space Telescopes at the Space Telescope Science Institute, which is operated by the Association of Universities for Research in Astronomy, Inc., under NASA contract NAS 5-03127 for JWST. These observations are associated with program JWST-ERS-1324 and JWST-DDT-2756. We acknowledge financial support from NASA through grant JWST-ERS-1324. 
KB, MT, BM, ND acknowledge support from the Australian Research Council Centre of Excellence for All Sky Astrophysics in 3 Dimensions (ASTRO 3D), through project number CE170100013. 
KG and TN acknowledge support from Australian Research Council Laureate Fellowship FL180100060. 
BM acknowledges support from Australian Government Research Training Program (RTP) Scholarships and the Jean E Laby Foundation.  
We acknowledge financial support through grants PRIN-MIUR 2017WSCC32 and 2020SKSTHZ. 
MB acknowledges support from the ERC Grant FIRSTLIGHT and from Slovenian national research agency ARRS through grants N1-0238 and P1-0188. 
CM acknowledges support by the VILLUM FONDEN under grant 37459 and the Carlsberg Foundation under grant CF22-1322. 
The Cosmic Dawn Center (DAWN) is funded by the Danish National Research Foundation under grant DNRF140. 
We acknowledge support from the INAF Large Grant 2022 ``Extragalactic Surveys with JWST”  (PI Pentericci). 
EV acknowledges support from the INAF GO Grant 2022 "The revolution is around the corner: JWST will probe globular cluster precursors and Population III stellar clusters at cosmic dawn". 
MC acknowledges support from INAF Minigrant ``Reionization and fundamental cosmology with high-redshift galaxies". 
PS acknowledges INAF Mini Grant 2022 ``The evolution of passive galaxies through cosmic time". 
DM acknowledges financial support from program HST-GO-17231, provided through a grant from the STScI under NASA contract NAS5-26555.
\item Authors' contributions: KB identified the emission lines from the NIRSpec data, led the overall data analysis activities, produced all figures, and was primarily responsible for writing the Methods section. MT provided advice on the data analysis and on its physical interpretation, carried out the comparison to theoretical modeling and contributed associated text in Methods, and was primarily responsible for writing the Abstract and Main sections. NL led the SED fitting and contributed associated text in Methods. AC led the clumping analysis and contributed associated text in Methods. BM led the comparison to hydrodynamical simulations and contributed associated text in Methods. GRB led the NIRSpec data reduction. ND led the Lyman Break modeling. LY led the light profile fitting from imaging data. TT led the GLASS/ERS survey conception, design and execution as the Principal Investigator of the program, and contributed advice on the paper preparation. TJ and AH contributed to physical interpretation of the absorption lines. AH, CM, TM, TN, and XW contributed to the NIRSpec data reduction and to the development of the NIRSpec pipeline. AF, EM, CM, DP contributed to the NIRSpec data reduction and to the development of the NIRCam pipeline. All authors contributed comments during the research activities and the manuscript preparation.     
\item Competing interests: The authors declare no competing interests. 
\end{itemize}

\clearpage

\end{document}